\documentclass[aps,prb,reprint,amssymb,superscriptaddress,showpacs]{revtex4-1}

\usepackage{rotating}
\usepackage{amsmath}
\usepackage{amssymb}
\usepackage{bm}
\usepackage{color}
\usepackage{comment}
\usepackage{graphicx}
\usepackage[active]{srcltx}
\usepackage[sort&compress]{natbib}
\usepackage{amssymb}
\usepackage[dvips]{epsfig}
\usepackage{float}
\usepackage{epsfig}
\usepackage{subfigure}
\usepackage{dsfont}
\usepackage[normalem]{ulem}
\newcommand{\ba}{\begin{align}}
\newcommand{\ea}{\end{align}}
\newcommand{\be}{\begin{equation}}
\newcommand{\ee}{\end{equation}}
\newcommand{\bse}{\begin{subequations}}
\newcommand{\ese}{\end{subequations}}
\newcommand{\bml}{\begin{multline}}
\newcommand{\eml}{\end{multline}}
\newcommand{\bea}{\begin{eqnarray}}
\newcommand{\eea}{\end{eqnarray}}

\definecolor{orange}{RGB}{255,127,0}
\definecolor{darkgreen}{RGB}{0,175,0}

\renewcommand{\(}{\left(}
\renewcommand{\)}{\right)}

\DeclareSymbolFont{matha}{OML}{txmi}{m}{it}
\DeclareMathSymbol{\varv}{\mathord}{matha}{118}

\def\g{\gamma}

\def\d{\delta}
\def\D{\Delta}
\def\e{\epsilon}
\def\ve{\varepsilon}
\def\h{\eta}

\def\l{\lambda}
\def\L{\Lambda}

\def\r{\rho}
\def\s{\sigma}
\def\S{\Sigma}

\def\x{\chi}
\def\w{\omega}
\def\W{\Omega}

\def\qq		{{\bf q}}

\def\zero	{{\bf 0}}
\def\one	{\tens{\mathds{1}}}
\def\kk		{{\bf k}}

\def\g{\gamma}

\def\d{\delta}
\def\D{\Delta}
\def\e{\epsilon}
\def\ve{\varepsilon}
\def\h{\eta}

\def\l{\lambda}
\def\L{\Lambda}

\def\r{\rho}
\def\s{\sigma}
\def\S{\Sigma}

\def\x{\chi}

\def\w{\omega}
\def\wplus{\omega^+}

\def\W{\Omega}

\def\et{\tilde{\epsilon}}

\def\nt{\tilde{n}}
\def\mt{\tilde{m}}

\def\bld{{\mathbf d}}

\def\blj{{\mathbf j}}
\def\blk{{\mathbf k}}

\def\blq{{\mathbf q}}
\def\blr{{\mathbf r}}

\def\blv{{\mathbf v}}

\def\blx{{\mathbf x}}

\def\blA{{\mathbf A}}

\def\blG{{\mathbf G}}

\def\blP{{\mathbf P}}

\def\blR{{\mathbf R}}

\def\ra{\rightarrow}

\def\bra{\langle}
\def\ket{\rangle}
\def\grad{\mbox{\boldmath $\nabla$}}

\def\KS{{\rm KS}}
\def\Exc{{\rm Exc}}
\def\IP{{\rm IP}}
\def\QP{{\rm QP}}
\newcommand*\tens[1]{\underline{\underline{#1}}}

\newcommand{\hR}{{\hat R}}

\def\1op{\hat{\mathbbm{1}}}
\def\1{\mathbbm{1}}

\newcommand{\cnrmlb} {Istituto di Struttura della Materia of the National Research Council, Via Salaria Km 29.3,
I-00016 Monterotondo Stazione, Italy}

\newcommand{\cnrstoulouse} {Laboratoire de Physique Th\'eorique, CNRS, IRSAMC, Universit\'e Toulouse III - Paul Sabatier, 118 Route de Narbonne, 31062 Toulouse Cedex, France}
\newcommand{\unitoulouse} {Laboratoire de Chimie et Physique Quantiques, IRSAMC, Universit\'e Toulouse III - Paul Sabatier, CNRS, 118 Route de Narbonne, 31062 Toulouse Cedex, France}

\newcommand{\cnrsmarsiglia} {CNRS/Aix-Marseille Universit\'e, Centre Interdisciplinaire de Nanoscience de Marseille UMR 7325 Campus de Luminy, 13288 Marseille cedex 9, France}
\newcommand{\qub}{School of Mathematics and Physics, Queen's University Belfast, Belfast BT7 1NN, Northern Ireland, UK}

\newcommand{\etsf} {European Theoretical Spectroscopy Facility (ETSF)}

\begin{document}
 
 \title{Optical properties of periodic systems within the current-current response framework: pitfalls and remedies}

\affiliation{\cnrmlb}
\affiliation{\unitoulouse}
\affiliation{\cnrsmarsiglia}
\affiliation{\qub}
\affiliation{\cnrstoulouse}
\affiliation{\etsf}

\author{Davide Sangalli}
\affiliation{\cnrmlb} 
\affiliation{\etsf} 

\author{J.~A.~Berger}
\affiliation{\unitoulouse} 
\affiliation{\etsf} 

\author{Claudio Attaccalite}
\affiliation{\cnrsmarsiglia} 
\affiliation{\etsf} 

\author{Myrta Gr\"uning}
\affiliation{\qub} 
\affiliation{\etsf} 

\author{Pina Romaniello}
\affiliation{\cnrstoulouse} 
\affiliation{\etsf}

\date{\today}
\begin{abstract}
  We compare the optical absorption of extended systems using the density-density and current-current linear response functions calculated 
  within many-body perturbation theory. The two approaches are formally equivalent for a finite momentum $\qq$ of the external perturbation. 
  At $\qq=\zero$, however,  the equivalence is maintained only if a small $q$ expansion of the density-density response function is used. 
  Moreover, in practical calculations this equivalence can be lost if one naively extends the strategies usually employed in the density-based approach 
  to the current-based approach. Specifically we discuss the use of a smearing parameter or of the quasiparticle lifetimes to describe the finite width of the spectral peaks and the inclusion of electron-hole interaction.
  In those instances we show that the incorrect definition of the velocity operator and the violation of the conductivity sum rule introduce unphysical features in the optical absorption spectra of three paradigmatic systems: silicon (semiconductor), copper (metal) and lithium fluoride (insulator). We then demonstrate how to correctly introduce lifetime effects and electron-hole interactions within the current-based approach.  
\end{abstract} 
\maketitle

%********************************************************************************
\section{Introduction}
%********************************************************************************

The electromagnetic linear response of solids can be measured experimentally by applying a small external
perturbation-or probe-which induces a small change in the sample material.
This change is the response of the material to the perturbing field and it can have both longitudinal and transverse components, 
depending on the experimental setup and the inhomogeneity of the system.
Well-known examples of such experiments are the measurements of absorption, electron energy loss, Kerr and Faraday effects, and dichroism.

From the theoretical point of view the linear response of a system 
to longitudinal fields can be obtained from both
the density-density response function $\x_{\r\r}$ and the current-current response function $\tens{\x}_{jj}$,
which is a tensor~\cite{Strinati1988}.
Instead, the response to transverse fields can only be described using $\tens{\x}_{jj}$.
This is due to the fact that the density determines the longitudinal current---through the continuity equation---
but not the transverse current. An important instance where transverse electric fields come into play is the recent
experimental progress on topological insulators~\cite{Checkelsky2012,Chang2013}. Therefore approaches based on $\tens{\x}_{jj}$ are more general than those based
on $\x_{\r\r}$.  On the other hand the current--density based approach is more susceptible to numerical issues and instabilities. For example the paramagnetic and diamagnetic contributions to $\tens{\x}_{jj}$ must be treated on equal
footing~\cite{Kootstra2000,Raimbault2015} otherwise divergencies can arise.

The differences between the two response functions are intrinsically related to the different gauge used to define 
the coupling of the external field with the electrons.
While transverse fields can be described only in terms of the transverse component of the
vector potential, for longitudinal fields there are two options.
One can use a scalar potential, which couples to the electron density
of the system, or the longitudinal component of the vector potential, which couples to the current density.
In the optical limit, i.e., in the limit where the momentum $\mathbf{q}$ carried
by the external perturbation is negligible, these two gauges are called  length and velocity gauge, respectively.
In this limit the distinction between longitudinal, or parallel to $\blq$, and transverse, or perpendicular to $\blq$,
fields vanishes.  However the optical limit is also the case where
many subtle differences between the two approaches arise, fundamentally because the density-based approach
is ill-defined at $\mathbf{q}=0$.

Various works in the literature have discussed and compared the two approaches either 
within the independent-particle (IP) picture~\cite{Sipe1993,Lamb1987,Sangalli2012}, 
in which optical properties can be described in terms of a simple sum-over-states expression,
or within a generalized Kohn-Sham approach.~\cite{Springborg2008} 
These studies focus mostly on the limitations of the density-based approaches and the avoidance of divergencies in the
current-based approaches.
An important point of debate is whether the two approaches give the same longitudinal macroscopic dielectric function. At the 
linear-response level the equivalence has been shown, but only in some specific cases, i.e., for semiconductors and insulators, 
for Hamiltonians with only local operators, and for absorption at resonance~\cite{Schafer}.  
The comparison for metals, for Hamiltonians with nonlocal operators, and for absorption out of resonance due to smearing,
~\footnote{
The differences out of resonance can be of small importance in the linear regime but they become crucial
for any non-linear phenomena because they play an important role in the construction
of high order response functions.} has received less or no attention.
More importantly,  no systematic comparison of the density-based and the current-based approaches has been carried out so far at the
many-body level.

The objectives of this work are \emph{(i)} to better elucidate the differences between the density and the current-based approaches at the IP level;
\emph{(ii)} to extend the discussion to the many-body framework. In particular, at the IP level, we compare the two approaches when lifetime effects are introduced (either from first-principles or by introducing a smearing parameter). At the many-body level we consider the Bethe-Salpeter equation (BSE). Based on Green functions theory, the BSE is the fundamental equation of Many-Body Perturbation Theory for the description of linear response properties~\cite{Fetter2003}.
In particular we address some important and rather subtle aspects related to the definition of the velocity operator in the current-based BSE approach~\cite{Louie2000}
which did not receive proper attention so far.~\cite{Strinati1988,Louie2000,DelSole1984}

We note that alternative approaches exist in which the explicit calculation of
response functions is avoided, for example by calculating the induced current-density in the
frequency domain~\cite{Kootstra2000,Romaniello2005,Berger2005,Berger2006}
or by calculating the time-dependent density or current-density using real-time
propagation~\cite{Attaccalite2011,Castro2006,Bertsch2000}. 
Many of the issues discussed in this work are also pertinent to these methods.

The manuscript is organized as follows. 
In Sec.~\ref{Sec:Background} we compare the density and the current-based approaches at a formal level, 
within both the independent-particle approximation and the BSE.
We compare the two approaches in the optical limit $\mathbf{q}\rightarrow \zero$ and elucidate the origin 
of their fundamental difference at $\mathbf{q}=\zero$.
In Sec~\ref{Sec:Results} we show how an incorrect implementation of smearing and an erroneous
definition of the velocity operator can lead to different spectra in the two approaches.
We then show how these problems can be solved. For sake of completeness, in Sec.~\ref{sec:Induced_quant}  we briefly discuss alternative approaches to calculate optical spectra. Finally, in Sec.~\ref{Sec:Conclusions} we draw our conclusions.

%********************************************************************************
\section{Formal equivalence of density and current-based approaches} \label{Sec:Background}
%********************************************************************************
In this section we review the basic equations which describe optical properties within both a density-based and a current-based approach
and we compare these approaches in the optical limit $\mathbf{q}\rightarrow \zero$. 
We first consider the case of noninteracting electron-hole pairs
referred to in the following as IP level,
\footnote{In condensed matter the IP approximations is often called RPA to be distinguished from RPA plus local field effects.}
and then the case in which the electron-hole interaction is treated within the BSE of Many-Body Perturbation Theory.
In both cases we demonstrate that the two approaches should lead to the same results for the longitudinal macroscopic dielectric function 
(see App.~\ref{App:trans_comp} for its definition).

\subsection{Noninteracting electron-hole pairs} \label{sec:basic}
%********************************************************************************
Let us consider a collection of electrons moving in a periodic potential ${v_0(\blx)=v_0(\blr)}$,
with $\blx=\blr+\blR$ and $\blr$ inside the unit cell volume $V$, whose position
is identified by the Bravais lattice vector $\blR$.
We are interested in the response of the system to a macroscopic time-dependent electromagnetic
field characterized by a set of scalar and
vector potentials, $\{\d\phi(\blx,t), \d\blA(\blx,t)\}$. A macroscopic quantity is defined as an average taken
over the unit cell whose location is given by $\blR$~\cite{Wiser1963} (see App.~\ref{App:FT_in_PBC}).

The motion of the system
is then governed by the following one-particle Hamiltonian,
\be
\hat{h}(t)=
\left\{\frac12[-i\nabla_\blx +\frac1c\mathbf{A}(\blx,t)]^2+v(\blx,t) \right\}
\label{Eqn:IP_H}
\ee
where $\blx=\blr+\blR$, and 
\begin{subequations}
\begin{align}
   v(\blx,t) &=v_0(\blr)+\d \phi(\blR,t), 
   \label{Eqn:PPa}\\
\blA(\blx,t) &=          \d \blA(\blR,t),
\label{Eqn:PPb}
\end{align}
\end{subequations}
for which the macroscopic external perturbations
$\delta\phi(\blR,t)$ and $\delta \blA(\blR,t)$
vanish identically for $t\leq 0$. The $v_0(\blr)$ term instead describes a periodic potential (i.e. it does not depend on $\blR$) and can be seen as the mean field felt by the electrons. Here and throughout the article we use atomic units $(\hbar=m_e=e=4\pi\ve_0=1)$ with the gaussian convention for electromagnetism. It is important to notice that the perturbing potentials $\delta\phi$ and $\delta\blA$ are external potentials and they do not take into account the contributions due to the response of the system. 
 
 The density and the current-density of the system are defined in terms
of the Bloch wave functions $\psi_n(\blx,t)$, which are eigenstates
of the one-particle hamiltonian $\hat{h}(t)$, as
\begin{subequations}
\label{Eqn:rho-j_def_eq}
\begin{align}
\r(\blx,t)  &= \sum_n f_n\, \psi^*_n(\blx,t)\, \hat{\r}\, \psi_n(\blx,t),      \label{Eqn:rho_def_eq} \\
\blj(\blx,t)&= \sum_n f_n\, \psi^*_n(\blx,t)\, \hat{\blv}(t)\, \psi_n(\blx,t). \label{Eqn:j_def_eq}
\end{align}
\end{subequations}
Here $n=\{\nt\kk\}$ is a generalized index that comprises the band index $\nt$
and the wave vector $\kk$, and $f_n$ are the occupations factors. In the collinear case $\nt=\{i\s\}$ can be further specified as
the band with the corresponding spin index $\s$, while in the spinorial case it is the 
spinorial band index.   
The density operator is the identity in real space $\hat{\r}=\hat{1}$.
The current-density is given in terms of the velocity operator~\cite{landau1965quantum}.
\bea
\hat {\mathbf{v}}(t)&\equiv&-i[\hat{h}_+(t),\blx]-i\{\hat{h}_-(t),\blx\}\label{Eqn:velocity_operator_general} 
\label{Eqn:velocity_operator}
\eea
where $\hat{h}_{\pm}=(\hat{h}\pm\hat{h}^\dagger)/2$, $[\hat{a},\hat{b}]$ is
the commutator, while $\{\hat{a},\hat{b}\}$ is the anticommutator.
With the Hermitian Hamiltonian defined in Eq.~\eqref{Eqn:IP_H} ${\hat {\mathbf{v}}(t)=(-i\nabla_\blx +\mathbf{A}(\blx,t)/c)}$.
The momentum operator $-i\nabla_\blx$ gives the paramagnetic
current density, i.e. ${\hat{j}^p=-i\nabla_\blx}$, and the potential-dependent term gives the diamagnetic current density. 
We note that the velocity operator, and therefore the current operator, depends on the Hamiltonian.
We will discuss this important point later in the paper.
Density and current-density variations
are induced as a response to the perturbing potentials in Eqs. (\ref{Eqn:PPa}) and (\ref{Eqn:PPb}). 

We restrict ourselves to macroscopic longitudinal perturbations
with a small transferred momentum $\blq$. 
In the following, therefore, we consider the longitudinali, i.e., parallel to $\blq$, component only
of the vector potential and the current (see App.~\ref{App:trans_comp}).
Longitudinal perturbations can be described by a scalar potential $\phi$,
for example in the Coulomb gauge $\grad \cdot\blA=0$, or
by a longitudinal vector potential $\blA$, for example in the (incomplete)
Weyl gauge $\delta\phi=0$. In the optical limit the two gauges generate the so-called
length and velocity gauges (see App.~\ref{App:gauges}).

The macroscopic density-density,and the longitudinal current-current response functions,  $\chi_{\rho\rho}$ and  $\chi_{jj}$, respectively, can be written in reciprocal space as (see App.~\ref{App:LR_C})
\begin{subequations}
\begin{align}
\chi_{\rho\rho}(\blq,\omega)&\equiv 
    \frac{\delta \rho(\blq,\omega)}{\delta \phi(\blq,\omega)},
 \label{Eqn:chi_rr_IP} \\
\chi_{jj}(\blq,\omega)      &\equiv 
   c\frac{\delta j(\blq,\omega)}{\d A(\blq,\omega)}
 \label{Eqn:chi_jj_IP}.
\end{align}
\label{Eqn:chi_IP}
\end{subequations}
When the induced density and current density are calculated from the Hamiltonian in (\ref{Eqn:IP_H}),
then the response functions in Eqs.~\eqref{Eqn:chi_IP} are the independent particle ones, $\chi^{\IP}_{\rho\rho}$ and $\chi^{\IP}_{jj}$.  
We can now define the longitudinal component of the macroscopic dielectric tensor $\ve(\qq; \w)$, 
which conveniently describes the optical properties of semiconductors in the long wavelength limit.  
It can be obtained, for $\w\neq 0$,
in terms of the longitudinal current-current response function as \cite{Strinati1988}
\be
\ve[\x^{\IP}_{jj}](\qq,\w)=1-\frac{4\pi}{\w^2}\x^{\IP}_{jj}(\qq,\w).
\label{Eqn:epsilon_jj_IP}
\ee 
For $\qq \neq \zero$, the dielectric function can also be expressed in terms of the density-density response function  as
\be
\ve[\x^{\IP}_{\r\r}](\qq,\w)=1-\frac{4\pi}{q^2}\x^{\IP}_{\r\r}(\qq,\omega),
\label{Eqn:epsilon_rr_IP}
\ee
with $q=|\qq|$.
Equation (\ref{Eqn:epsilon_rr_IP}), in the limit $\qq \rightarrow \zero$, is the expression commonly used for the calculations of optical properties in solids.
For both expressions (\ref{Eqn:epsilon_jj_IP}) and (\ref{Eqn:epsilon_rr_IP}) the non-analytic point $(\qq,\w)=(\zero,0)$ can be described only 
via a limiting procedure and the final result depends on the direction of the limit.

Both $\ve[\x_{jj}]$ and $\ve[\x_{\r\r}]$ should lead to the same result.
Indeed the two response functions are related by the expression~\cite{Giuliani,Schafer}
\be
q^2\, \x_{jj}(\qq,\w)=\w^2\, \x_{\r\r}(\qq,\w),
\label{Eqn:Xjj_Xrr_relation}
\ee
which follows from the continuity equation ${\qq\cdot\mathbf{j}=\w \r}$,
which guarantees local charge conservation.
Thus at ${\qq\neq\zero}$ and ${\w\neq 0}$ relation (\ref{Eqn:Xjj_Xrr_relation}) ensures
$\ve[\x_{jj}]=\ve[\x_{\r\r}]$.
We note that although we have established formal equivalence this does not guarantee numerical equivalence,
as we show in Section~\ref{Sec:Results}.

\subsubsection{Choice of the reference one-particle hamiltonian} \label{Sec:IP}
%********************************************************************************
Within the IP picture defined by the Hamiltonian in Eq. (\ref{Eqn:IP_H}) the response functions can be written as
\be
\x^{\IP}_{aa}(\blq,\omega)=
 -\frac{1}{V}\sum_{nm}
   \frac{a^{\IP}_{\qq,(nm)}\, \tilde{a}^{\IP}_{\qq,(mn)}}
       {\wplus-\D\e^{\IP}_{mn}(\blq)}
\label{Eqn:chi_0}
\ee
with 
\bea
\D\e^{\IP}_{mn}(\blq)&=&\e^{\IP}_{\mt\blk+\blq}-\e^{\IP}_{\nt\blk}                           \\
\D f_{mn}(\blq)     &=&f_{\mt\blk+\blq}-f_{\nt\blk}                                       \\
a^{\IP}_{\qq,(nm)}   &=&\langle \nt\kk|e^{-i\blq\cdot\blr} \hat{a}|\mt\kk+\qq\rangle
                      \,\sqrt{\D f_{mn}(\blq)}\label{Eqn:IP_a}     \\       
\tilde{a}^{\IP}_{\qq,(mn)}   &=&\langle\mt\kk+\qq|e^{i\blq\cdot\blr} \hat{a}| \nt\kk\rangle\,\sqrt{\D f_{mn}(\qq)}
                                \label{Eqn:IP_operators},
\eea
(the spin index in $n$ and $m$ is identical in the collinear case) and 
\be
\wplus =\lim_{\eta\ra 0} (\w+i\eta),          \label{Eqn:wplus_def}    
\ee
where it is understood that the limit $\eta\rightarrow 0$ is taken at the end of the calculation.
Here $\hat{a}$ is either the one-particle density operator $\hat{\rho}$ or
the paramagnetic current-density operator $\hat{j}^p$, and $|n\rangle$ are Bloch states, which are eigenstates of the equilibrium one-particle Hamiltonian $\hat{h}(t=0$), 
with corresponding energies $\e^{\IP}_n$ and occupation numbers ${0\leq f_n\leq 1}$.\footnote{The inclusion of $\sqrt{\D f_{mn}(\qq)}$ in the definition of the matrix elements
of Eq.~\eqref{Eqn:IP_operators} allows to define in the following sections (where the 
the case of interacting electron-hole pairs is considered) an excitonic matrix
which remains Hermitian also in the general case of fractional occupation numbers.
See also Ref.~\onlinecite{Sangalli2016}.}
It is important to notice that the transitions with ${\D\e^{\IP}_{mn}(\blq)=0}$
give no contribution since at equilibrium $\Delta{f_{mn}(\blq)=0}$. 
Thus all the summations in the present manuscript are intended without the zero energy poles.
The full current-current response function is obtained via
\be
\x^{\IP}_{jj}(\blq,\omega)=\x^{\IP}_{j^pj^p}(\blq,\omega)+\frac{N}{V}. 
\label{Eqn:CCD}
\ee
We shall now choose the stationary part of the one-particle Hamiltonian (\ref{Eqn:IP_H}).
A  common choice is the Kohn-Sham Hamiltonian from density functional theory (DFT) where
\be
v_0(\blr)=v_N(\blr)+v_H(\blr)+v_{xc}(\blr),
\ee
is the equilibrium Kohn-Sham (KS) potential.
It is the sum of the potential generated by the nuclei, $v_N(\blr)$, the Hartree potential,
$v_H(\blr)$, and the exchange-correlation potential, $v_{xc}(\blr)$.
We refer to the IP response function derived from the KS hamiltonian as $\chi^{\KS}_{aa}$.
However, usually DFT-KS band structures are not a good starting point for response calculations, since, for example, the fundamental band gap is
systematically underestimated in semiconductors and insulators. To improve over DFT one can introduce 
a more general nonlocal and frequency dependent quasiparticle (QP) potential
\be
v_{0,\sigma\sigma'}(\blx,\blx',\w)=v_N(\blr)+v_H(\blr)+\S_{xc,\sigma\sigma'}(\blx,\blx',\w),
\ee
where $\S_{xc}$ is the exchange-correlation many-body self-energy, defined within
the formalism of Many-Body Perturbation Theory (MBPT), evaluated at
equilibrium.
In particular we consider the (first-order) QP hamiltonian $H^{\QP}$ 
defined as the first-order correction in the perturbation
${\S_{xc} - v_{xc}}$ of the KS hamiltonian:
\be
H^{\QP}_{nm}(\w)=\delta_{n,m}\Big[ \e^{\KS}_{n}+ \bra n| \S_{xc}(\w) - v_{xc}| n \ket\Big].
\label{Eqn:QPH}
\ee
We have here emphasized the $\w$ dependence of $\S_{xc}$, 
and dropped the $\blr$ space and $\s$ spin dependence of both $\S_{xc}$ and $v_{xc}$.
The QP eigenvalues are thus corrected KS energies
\be
\e^{\QP}_{n}=\e^{\KS}_{n}+\langle n |\, \Sigma_{xc}\!\left(\e^{\QP}_{n}\right)-v_{xc}\, |n\rangle,
\label{Eqn:e_QP}
\ee 
while we keep the same wave functions ${\psi^{\QP}_n=\psi^{\KS}_n}$.
From the quasiparticle energies $\e^{\QP}$, solution of Eq.~(\ref{Eqn:e_QP}), we can define the QP
response function, $\x^{\QP}_{aa}$, which differs from the KS response function 
by the replacement $\e^{\KS}_n\ra\e^{\QP}_{n}$.

\subsubsection{Expansion of the current-based approach at finite momentum} \label{Sec:Xjj_expansion}
%********************************************************************************
We now focus on the relation between the density-based and current-based formalism by 
extending the approach of Ref.~\onlinecite{Sipe1993} to $\qq\neq\mathbf{0}$.
To simplify the notation we rewrite the response function (\ref{Eqn:chi_0}) as 
\be
\x^{\IP}_{aa}(\blq,\omega)=-
 \sum_{nm}
   \frac{K^{aa}_{nm,\qq}}
       {\wplus-\D\e^{\IP}_{mn}(\blq)},
\label{Eqn:general_chi}
\ee
where $K^{aa}_{nm,\qq}=a^{\IP}_{\qq,(nm)}\, \tilde{a}^{\IP}_{\qq,(mn)}/V$.
By using the exact relation~\footnote{which can be obtained by Taylor expanding ${F(\w)}$ around ${\w=0}$ up to the first order.
The last term on the right-hand side is then defined from the difference ${F(\w)-F(0)-F'(0)\,\w}$.}
\be
F(\w)=\frac{1}{\w-\Delta\epsilon}=-\frac{1}{ \Delta\epsilon} -\frac{\w}{ \Delta\epsilon^2}
 + \frac{\w^2}{(\w-\Delta\epsilon)\Delta\epsilon^2},
\label{Eqn:exact_TS}
\ee
in Eq.\eqref{Eqn:general_chi} and inserting the result into Eq.\eqref{Eqn:epsilon_jj_IP},
we can decompose $\ve[\x^{\IP}_{jj}](\qq,\w)$ as
\be
\ve[\x^{\IP}_{jj}](\qq,\w)=1+\frac{A^{\IP}(\qq)}{\w^2}+\frac{B^{\IP}(\qq)}{\w}+C^{\IP}(\qq,\omega),
\label{Eqn:eps_jj_expansion}
\ee
with 
\begin{subequations}
\label{Eqn:ABC_expr}
\begin{align}
A^{\IP}(\qq)    &=   -4\pi\left[ \sum_{nm}\frac{K^{j^pj^p}_{nm,\qq}}{\Delta\epsilon_{nm}(\qq)}+\frac{N}{V}\right]
= -4\pi  \x^{\IP}_{jj}(\qq,0), \label{Eqn:A_expr} \\
B^{\IP}(\qq)    &=   -4\pi \sum_{nm}\frac{K^{j^pj^p}_{nm,\qq}}{\Delta\epsilon_{nm}^2(\qq)} , \label{Eqn:B_expr}                 \\
C^{\IP}(\qq,\w) &=\ \ 4\pi \sum_{nm}\frac{K^{j^pj^p}_{nm,\qq}}{\Delta\epsilon_{nm}^2(\qq)(\w-\Delta\epsilon_{nm}(\qq))}, \label{Eqn:C_expr}
\end{align}
\end{subequations}
where in Eq.\eqref{Eqn:A_expr} we used Eqs.\eqref{Eqn:CCD} and \eqref{Eqn:general_chi}.
From the conductivity sum rule (CSR) for $\x^{\IP}_{jj}$, given by
\be
\x^{\IP}_{jj}(\qq,0)=\x^{\IP}_{j^pj^p}(\qq,0)+\frac{N}{V}=0,
 \label{Eqn:CSR}
\ee
we see that $A^{\IP}(\qq)=0$.
Moreover, if time-reversal symmetry holds, also $B^{\IP}(\qq)=0$.
Therefore, in case of time-reversal symmetry, only the $C(\qq,\w)$ term survives and we have,
at finite momentum, the general result
\be
\ve[\x^{\IP}_{jj}](\qq,\w)=1+C^{\IP}(\qq,\w)=\ve[\x^{\IP}_{\r\r}](\qq,\w).
\label{Eqn:eps_jj_rr_relation}
\ee 
The last equality~\footnote{Eq.~\eqref{Eqn:eps_jj_rr_relation} implies 
$q^2 K^{j^pj^p}_{nm,\qq}=\D\e_{nm}^2(\qq) K^{\r\r}_{nm,\qq}$, which is similar, but not
the same, to Eq.~\eqref{Eqn:Xjj_Xrr_relation}. Notice the non trivial replacement of $\w^2\ra\D\e_{nm}^2(\qq)$.}
holds thanks to Eq.~\eqref{Eqn:Xjj_Xrr_relation}. 
In the next section we will use (\ref{Eqn:A_expr})-(\ref{Eqn:C_expr}) to discuss the optical limit.

\subsubsection{The optical limit}  \label{Sec:Optical_limit}
%********************************************************************************
%
We are interested in computing optical properties, for which $\w=c\,q$, with $\omega$ in the order of a few eV.
Therefore we consider $q\approx 0$ compared to the size of the Brillouin zone. 
However, $\ve[\x^{\IP}_{\r\r}]$, given in Eq.~(\ref{Eqn:epsilon_rr_IP}), is not defined for $q=0$ because of the $1/q^2$ term.
Therefore, to obtain an explicit expression for $\ve[\x^{\IP}_{\r\r}]$ for small $q$,
we Taylor expand $\r^{\IP}_{\qq,(nm)}$, defined in Eq.~(\ref{Eqn:IP_operators}), around $q=0$,
\be
\r^{\IP}_{\qq,(nm)}= -i\,q\,d^{\IP}_{\qq,(nm)}+O(q^2),
\ee
where~\footnote{The position operator is ill-defined when periodic boundary conditions (PBC) are imposed. 
Here we implicitly work in the crystal momentum representation in which the matrix elements of the position operator
are redefined consistently with the PBC~\cite{Blount1962,Souza2004}}
${d^{\IP}_{\qq,(nm)}=\langle n|\hat\qq\cdot \blx|m\rangle\sqrt{\D f_{nm}(\qq)}}$ with $\hat\qq=\qq/q$ the direction of $\qq$.
Substitution into Eq.~(\ref{Eqn:general_chi}) leads to 
$\x^{\IP}_{\rho\rho}(\blq,\omega)=q^2\x^{\IP}_{dd}(\blq,\omega) + O(q^3)$
with
\bea
\x^{\IP}_{dd}(\blq,\omega)=
  -&&\sum_{nm}
   \frac{K^{dd}_{nm,\qq}}
       {\wplus-\D\e^{\IP}_{mn}(\blq)}
\label{Eqn:chi_dd}
\eea
the longitudinal dipole-dipole response function.
Thus, for small $q$, we can rewrite Eq.~(\ref{Eqn:epsilon_rr_IP}) as
\be
\ve[\x^{\IP}_{dd}](\qq,\w)=  1 -4\pi\sum_{nm}
   \frac{K^{dd}_{nm,\qq}}
       {\wplus-\D\e^{\IP}_{mn}(\blq)},
\label{Eqn:eps_dd}
\ee
which is well defined at $\qq=\zero$.\footnote{Here $d_{\blq}$ at $\qq=\zero$ is the longitudinal dipole,
defined by the direction of the $\qq\rightarrow \zero$. See also App.~\ref{App:trans_comp}.} 
It is Eq.~(\ref{Eqn:eps_dd}) that is usually implemented
to compute absorption spectra of cold semiconductors and insulators in the density-based approach.

However, Eq.~\eqref{Eqn:eps_dd} is formally exact only in the limit $\qq\rightarrow\zero$,
since at $\qq = \zero$ contributions from intraband transitions, i.e., transitions within a single band, are excluded
\footnote{For finite $\qq$ Eq.~\eqref{Eqn:eps_dd} is exact to first order in $\qq$}.
Instead, no such problem exists for the current-based approach.
We can summarize this difference between the two approaches as
\begin{subequations}
\label{Eqn:X_limit}
\begin{align}
\lim_{\qq\ra\zero}\ve[\x^{\IP}_{jj}](\qq,\w)  &  = \ve[\x^{\IP}_{jj}](\zero,\w), \label{Eqn:X_jj_limit}\\
\lim_{\qq\ra\zero}\ve[\x^{\IP}_{dd}](\qq,\w) &\neq \ve[\x^{\IP}_{dd}](\zero,\w) . \label{Eqn:X_rr_limit}
\end{align}
\end{subequations}
We now discuss the cases $\qq=\zero$ and $\qq\rightarrow\zero$ in more detail by separating $A^{\IP}(\qq)$, $B^{\IP}(\qq)$ and $C^{\IP}(\qq)$
into inter- and intraband contributions. Moreover, $A^{\IP}(\qq)$ also contains the constant diamagnetic term $A^d=-4\pi N/V$.

\underline{\emph{The $\qq=\zero$ case: interband transitions.}}
\newline
At ${\qq=\zero}$ only interband transitions,
i.e., transitions between two different bands, contribute to the summations
in Eqs.~(\ref{Eqn:A_expr})-(\ref{Eqn:C_expr}), since for intraband transitions $\Delta f_{nn}=0$.
Because of the missing intraband contributions in $\x^{\IP}_{jj}$ the CSR in Eq.~(\ref{Eqn:CSR}) in general does not hold at $\qq=\zero$. 
Let us define
\begin{subequations}
\label{Eqn:interband_def}
\begin{align}
A^\textrm{IP,inter}    &\equiv A^{\IP}(\zero)-A^d, \label{Eqn:Ae_def} \\
B^\textrm{IP,inter}    &\equiv B^{\IP}(\zero),     \label{Eqn:Be_def} \\ 
C^\textrm{IP,inter}(\w)&\equiv C^{\IP}(\zero,\w).  \label{Eqn:Ce_def}
\end{align}
\end{subequations}
One can verify~\footnote{Here $B^\textrm{IP,inter}=0$ because we are considering the longitudinal term only.
Indeed the mixed longitudinal-transverse terms can instead be different from zero
and describe the Anomalous Hall effect~\cite{Sangalli2012}.}
that $B^{\IP}(\zero)=0$ by exchanging $n$ and $m$ in Eq.~\eqref{Eqn:B_expr}.

Let us first consider systems with a gap. 
Then $A^\textrm{IP,inter}+A^d=0$ since the dielectric function
must go to a constant~\cite{Sipe1993} as $\w\ra 0$.
We can thus focus on $C^\textrm{IP,inter}(\w)$.
Using Eq.~(\ref{Eqn:IP_a}) for the paramagnetic current-density, we can write
\bea
j^{p,IP}_{\zero,(nm)}&=& \langle n | \hat{\mathbf{v}} |  m \rangle  \,\sqrt{\D f_{nm}(\zero)}, \nonumber \\
              &=&  -i\,\langle n | \hat{\mathbf{x}} | m \rangle \,\D\e_{nm}(\zero)\,\sqrt{\D f_{nm}(\zero)}, \nonumber \\
              &=& -i\,d^{\IP}_{\zero,(nm)}\,\D\e_{nm}(\zero),
\label{Eqn:IP_vel_operator}
\eea
where we used $\hat{\mathbf{v}}=-i[\hat{h},\hat{\mathbf{x}}]$ and $d^{\IP}_{\zero,(nm)}$ given by Eq.~(\ref{Eqn:IP_a}) with $\hat{a}=\hat{\mathbf{x}}$.
From this relation we deduce that
\be
K^{dd}_{nm,\zero} = \frac{K^{j^pj^p}_{nm,\zero}}{\Delta\epsilon^2_{nm}(\zero)}.
\label{Eqn:Kjj-Kdd}
\ee
Substitution of this identity into Eq.~(\ref{Eqn:eps_dd}) shows that at $\qq=\zero$
both $\ve[\x^{\IP}_{jj}]$ and $\ve[\x^{\IP}_{dd}]$
can be expressed in terms of $C^\textrm{IP,inter}$
for systems with a gap:
\be
\ve[\x^{\IP}_{jj}](\zero,\w)=\ve[\x^{\IP}_{dd}](\zero,\w)=1+C^\textrm{IP,inter}(\w).
\label{Eqn:Xrr_Xjj_identity_gap_q0}
\ee
Equation \eqref{Eqn:Xrr_Xjj_identity_gap_q0} proves the equivalence between the density-based and
the current-based approaches for cold semiconductors and insulators.

For metals, however, we also need to describe the divergent Drude-like term.
Since the current-based approach is exact also at $\qq=\zero$, this term must be described
by $(A^\textrm{IP,inter}+A_d)/\w^2$, i.e. the Drude tail originates from the breaking of the CSR. 
Thus we have
\be
\ve[\x^{\IP}_{jj}](\zero,\w)=1+\frac{A^\textrm{IP,inter}+A^d}{\w^2}+C^\textrm{IP,inter}(\w).
\label{Eqn:current_full}
\ee
Instead, the density-based approach does not
contain any extra term beyond $C^\textrm{IP,inter}(\w)$ and thus cannot describe metals at $\qq=\zero$.
The Drude-like tails in the density-based approach can be obtained only explicitly dealing with the $\qq\ra\zero$ limit
as explained in the next subsection.

\underline{\emph{The $\qq\ra\zero$ limit: intraband transitions.}}
\newline
In the $\qq\ra\zero$ limit also intraband transitions contribute to the summations
in Eqs.~(\ref{Eqn:A_expr})-(\ref{Eqn:C_expr}).
One can thus define
\begin{subequations}
\label{Eqn:intraband_def}
\begin{align}
&A^\textrm{IP,intra}=\lim_{\qq\ra\zero}A^{\IP}(\qq)-A^{\IP}(\zero)=-A^{\IP}(\zero),                \label{Eqn:Ai_def} \\
&B^\textrm{IP,intra}=\lim_{\qq\ra\zero}B^{\IP}(\qq)-B^{\IP}(\zero)=\lim_{\qq\ra\zero}B^{\IP}(\qq), \label{Eqn:Bi_def} \\
&C^\textrm{IP,intra}(\w)=\lim_{\qq\ra\zero}C^{\IP}(\qq,\w)-C^{\IP}(\zero,\w)                 \label{Eqn:Ci_def},
\end{align}
\end{subequations}
where we used Eq.~(\ref{Eqn:Be_def}) and the fact that $A^{\IP}(\qq\neq\zero)=\zero$ owing to the CSR in Eq.~\eqref{Eqn:CSR}.
If time-reversal symmetry holds $B^{\IP}(\qq)=0$ and hence $B^\textrm{IP,intra}=0$.
We note that here and in the rest of the paper the $\qq\rightarrow 0$ limit is taken at finite $\omega$.
Using the results of Eqs.~(\ref{Eqn:Ai_def})-(\ref{Eqn:Ci_def}) in Eq.~(\ref{Eqn:eps_jj_expansion}) we obtain
\be
\lim_{\qq\ra\zero}\ve[\x^{\IP}_{jj}](\qq,\w)=1+C^\textrm{IP,intra}(\w)+C^\textrm{IP,inter}(\w)
\ee
where we used Eq.~(\ref{Eqn:Ce_def}).
Owing to Eq.~(\ref{Eqn:X_jj_limit}) we can compare this result to Eq.~(\ref{Eqn:current_full}) and conclude that
\be
C^\textrm{IP,intra}(\omega)=\frac{A^\textrm{IP,inter}+A^d}{\omega^2}.
\label{Eqn:Ci_drude}
\ee
This means that, within the current-based approach, the Drude-like tail, which at $\qq=\zero$
is described by $(A^\textrm{IP,inter}+A^d)/\omega^2$, in the limit $\qq\ra\zero$ is described via $C^\textrm{IP,intra}(\omega)$.
Indeed, the exact expression for $\ve[\x^{\IP}_{jj}]$ at finite $\qq$ in Eq.~\eqref{Eqn:eps_jj_rr_relation} 
only depends on $C^{\IP}(\qq,\w)$.
Thus $C^\textrm{IP,intra}(\w)$ must describe all intraband transitions in metals when $\qq\rightarrow\zero$.\cite{MariniPhD}.

To have an explicit expression of such intraband contribution in the
density-based approach, one needs to Taylor expand the
energy and occupation number differences as~\cite{Romaniello2005}
\bea
(\epsilon_{n\mathbf{k}}-\epsilon_{n\mathbf{k}+\qq})&=&- \varv_{n\kk}\cdot\qq+O(q^2)
\label{Eqn:intraband-energies}\\
(f_{n\mathbf{k}}-f_{n\mathbf{k}+\qq})&=&-q\frac{df}{d\epsilon}(\nabla_\kk\ \epsilon_{n\kk}\cdot\hat{\qq})+O(q^2),
\label{Eqn:intraband-energies_2}
\eea
where $\varv_{n\kk}=\nabla_\kk\epsilon_{n\kk}$. At the IP level there is no Drude tail in the absorption (see App.~ \ref{App:intra} for details). Only when introducing a  smearing a Drude-like peak in the absorption appears~\cite{Romaniello2005,Berger2005}.
Experimentally such smearing also exist, because of the interaction between electrons
and thus in practice a peak is always measured.
 
In conclusion, the current-based approach is exact both in the limit $\qq\rightarrow \zero$
and at $\qq=\zero$. In particular, if metals are considered, the Drude-like tail is present in both
cases, but it is described by different terms, namely by $C^{\textrm{IP,intra}}(\omega)$ in
the limit $\qq\rightarrow \zero$ and by $(A^{\textrm{IP,inter}}+A^d)/\omega^2$ at $\qq=\zero$.
Also the density-based approach is exact in the limit $\qq\rightarrow \zero$. 
This can be summarized by the following set of relations
\bea
&&\ve[\x^{\IP}_{jj}](\zero,\w)=1+\frac{A^\textrm{IP,inter}+A^d}{\w^2}+C^\textrm{IP,inter}(\w) \nonumber\\
&=&\lim_{\qq\ra\zero}\ve[\x^{\IP}_{jj}](\qq,\w)=1+C^\textrm{IP,intra}(\w)+C^\textrm{IP,inter}(\w)\nonumber\\
&=&  \lim_{\qq\ra\zero}\ve[\x^{\IP}_{\r\r}](\qq,\w).
\nonumber
\eea

\subsection{Interacting electron-hole pairs}  \label{Sec:BSE}
%********************************************************************************
So far we have only considered independent particles.
We now also take into account the electron--hole interaction by considering the variations induced in the
Hartree and the exchange--correlation self--energy by the external potential. 
Within MBPT such variations can be conveniently
described using the BSE\cite{Strinati1988} for the (time-ordered) two-particle propagator $L$, which reads
\begin{multline}
\bar{L}(1,2,1^\prime,2^\prime)=L_0(1,2,1^\prime,2^\prime)+\int d3d4d5d6 \\
\times L_0(1,4,1^\prime,3)\bar{\Xi}(3,5,4,6)\bar{L}(6,2,5,2^\prime),
\label{Eqn:BSE}
\end{multline}
where 
$i\equiv {\blr_i,t_i}$.
In Eq.~\eqref{Eqn:BSE} $L_0$ is given in terms of QP Green's functions as
${L_0(1,2,1',2')=-iG^{\QP}(1,2')G^{\QP}(2,1^\prime)}$
and the four-point kernel $\bar{\Xi}(3,5,4,6)$ is given by
\bea
\bar{\Xi}(3,5,4,6)=i\frac{\delta\left[\bar{v}_H(3)\delta(3,4)+\delta\Sigma_{xc}(3,4)\right]}{\delta G(6,5)},
\eea
where $\bar{v}_H$ is the Hartree potential without the long-range component of the Coulomb potential, $v_c(\mathbf{G}=0)$. 
We are now looking at the self-consistent response of the system to a macroscopic field composed of both the external and the induced macroscopic field.
Therefore 
$v_c(\mathbf{G}=0)$ is effectively removed~\cite{Bussi2004}.
We use the $GW$ self-energy with an instantaneous screened Coulomb potential $W$.
In this case  the propagator $L$ depends only on the time difference
$\tau=t_1-t_2$\footnote{As commonly done in the literature, we neglect the term $i\delta W/\delta G$ in the kernel.}.
A Fourier transformation to frequency space leads to
\bea
\bar{L}(\omega)
&=&L_0(\omega)+L_0(\omega)\, \bar{\Xi}\, \bar{L}(\omega).
\label{Eqn:L(w_static)}
\eea
where, for notational convenience, we dropped the spin and space
arguments~\footnote{for a detailed treatment of the space indexes at finite momentum
see Ref.~[\onlinecite{Sottile2013}]; for the treatment of occupations factors see
Ref.~[\onlinecite{Sangalli2016}])}.

The BSE can be mapped onto an effective two-particle equation,
written in a basis of electron-hole transitions $(nm)$,
according to ~\cite{Onida2002,Sottile2013}
\begin{equation}
 H^{Exc}_{(nm)(n'm')}(\qq)\,
   A^{(n'm')}_{\lambda,\qq}=
     E_\lambda(\qq) A^{(nm)}_{\lambda,\qq},
 \label{Eqn:H2p_eigen}
\end{equation}
where $H^{Exc}(\qq)$ is the excitonic Hamiltonian with eigenvectors $A_{\lambda,\qq}$ and eigenvalues $E_\lambda(\qq)$.
The (retarded)\footnote{The exact Dyson equations holds only for the time-ordered two-particles propagator,
which is formally derived assuming zero temperature (i.e. integer occupation numbers
also in case of metals). To consider fractional occupations one would need to introduce
a finite temperature formalism. However, using a static kernel, the Dyson equation
at finite temperature reduces to a Dyson equation identical to Eq.~\eqref{Eqn:L(w_static)}
but for the retarded propagator, which is indeed what is needed to define
the dielectric function. Thus from now on we can consider all quantities as retarded functions
and forget about the time-ordered formalism.} particle-hole propagator $\bar{L}$
can be obtained \textit{via} 
 \bea
\bar{ L}_{(nm)(n'm')}(\qq,\w)=
  \sum_{\lambda\lambda'} 
    \frac{B^{(nm)}_{\lambda,\qq} S^{-1}_{\lambda\lambda',\qq}\, B^{*(n'm')}_{\lambda',\qq} }
      {\wplus-E_{\l}(\qq)} \nonumber, 
 \eea
where $B_{\lambda,\qq}$ and the overlap matrix $S_{\lambda\lambda^\prime,\qq}$ are defined by
\bea
B^{(nm)}_{\lambda,\qq}&=& A^{(nm)}_{\lambda,\qq} \sqrt{\Delta f_{mn}(\qq)}, \\ 
S_{\lambda\lambda^\prime,\qq}&=&\sum_{nm}A^{*(nm)}_{\lambda,\qq}A^{(nm)}_{\lambda^\prime,\qq}.
\eea

Upon solving Eq.~\eqref{Eqn:H2p_eigen}, one can obtain the density-density
and the current-current excitonic response functions $\bar{\chi}^{Exc}_{aa}$ as
\bea
\bar{\chi}^{Exc}_{aa}(\blq,\omega)=
  -\frac{1}{V}&&\sum_{\l\l'}\,
    a^{Exc}_{\qq,\lambda}\,\frac{S^{-1}_{\l\l',\qq}}{\wplus-E_\l(\qq)}\, \tilde{a}^{Exc}_{\qq,\l'}, 
\label{Eqn:chi_bse}
\eea
where $a^{Exc}_{\qq,\l}$ and $\tilde{a}^{Exc}_{\qq,\l'}$ are defined analogously to $a_{\qq,\l}$ and $\tilde{a}_{\qq,\l'}$ in
Eqs.~(\ref{Eqn:IP_a}) and (\ref{Eqn:IP_operators}) but the expectation value is with respect to the excitonic wave function.
As for the IP case, the zero-energy transitions give no contribution to the summation~\footnote{We assume that zero energy
transitions at the BSE level originates from zero energy transitions at the IP 
level. Zero energy poles at the BSE level may originate
from finite energies transition at the IP level as well. However this case would point
to an instability of the ground state which would be degenerate to an excited state.
We exclude this possibility in the present work.} 
since $\D f_{nm}(\qq)=0$ and thus ${B^{(nm)}_{\lambda,\qq}=0}$. 
Note that neglecting the kernel $\bar{\Xi}$ in the BSE \eqref{Eqn:L(w_static)},
Eq.~\eqref{Eqn:chi_bse} reduces to the $\textit{\IP}$ response function given in Eq.~(\ref{Eqn:chi_0}).

Substitution of $\bar{\x}^{Exc}_{jj}$ and $\bar{\x}^{Exc}_{\r\r}$ in Eqs.~(\ref{Eqn:epsilon_jj_IP}) and (\ref{Eqn:epsilon_rr_IP}), respectively, 
yields an expression for the macroscopic dielectric tensor in the current- and density-based approach.
As in the IP case the two expressions are identical for ${\qq\neq\zero}$ and ${\w\neq 0}$ thanks
to Eq.~\eqref{Eqn:Xjj_Xrr_relation}.

\subsubsection{The optical limit}\label{Sec:INTRA-BSE}
%********************************************************************************
The analysis for the optical limit in
Sec.~\ref{Sec:Xjj_expansion}-Sec.~\ref{Sec:Optical_limit} also applies to the excitonic case.
In particular, by rewriting the expression for the excitonic response function as
\be
\bar\x^{Exc}_{aa}(\qq,\w)=-\sum_{\l\l'}\frac{K^{aa}_{\l\l',\qq}}{\wplus-E_\l(\qq)},
\label{Eqn:general_chi_exc}
\ee
where $K^{aa}_{\l\l',\qq}=a_{\qq,\lambda}\,S^{-1}_{\l\l',\qq}\, \tilde{a}_{\qq,\l'}/V$
one obtains
\begin{subequations}
\label{Eqn:ABC_expr_BSE}
\begin{align}
A^{\Exc}(\qq)    &=   -4\pi\left[ \sum_{\l\l'}\frac{K^{j^pj^p}_{\l\l',\qq}}{E_\lambda(\qq)}+\frac{N}{V}\right], \label{Eqn:A_expr_BSE} \\
B^{\Exc}(\qq)    &=   -4\pi \sum_{\l\l'}\frac{K^{j^pj^p}_{\l\l',\qq}}{E_\lambda^2(\qq)} ,                       \label{Eqn:B_expr_BSE} \\
C^{\Exc}(\qq,\w) &=\ \ 4\pi \sum_{\l\l'}\frac{K^{j^pj^p}_{\l\l',\qq}}{E_\lambda^2(\qq)(\w-E_\lambda(\qq))}.     \label{Eqn:C_expr_BSE}
\end{align}
\end{subequations}
The interband dielectric function in terms of the excitonic dipole-dipole
response function $\chi^{\Exc}_{dd}$ reads
\be
\ve[\x^{\Exc}_{dd}](\qq,\w)=  1 +\frac{4\pi}{V}\sum_{\l\l'}
\frac{K^{dd}_{\lambda\lambda',\qq} }{\wplus-E_\l(\qq)},
\label{Eqn:eps_dd_exc}
\ee
where now $K^{dd}_{\lambda\lambda',\qq}$ contains the longitudinal excitonic dipoles
$d^{\Exc}_{\qq,\lambda}=\sum_{nm}A^{(nm)}_{\lambda,\qq}x_{(nm)}\,\sqrt{\D f_{nm}(\qq)}$. 

Finally, the analogous of the relation \eqref{Eqn:Kjj-Kdd} exists also for the exitonic case:
\be
\frac{K^{j^pj^p}_{\l\l',\zero}}{E^2_{\l}(\zero)}= K^{dd}_{\l\l',\zero},
\ee
from which expressions analogous to  Eqs. (34) and (35) can be written for $\ve[\chi^{\Exc}_{jj}]$.

%********************************************************************************
\section{Issues in the practical implementation of current-based approaches} \label{Sec:Results}
%********************************************************************************
In the previous section the equivalence between the current-based and density-based approaches was established 
for the longitudinal macroscopic dielectric function.
In the present section we compute optical absorption in a semiconductor (bulk silicon), an insulator (LiF), and a metal (copper)
comparing the two approaches numerically, i.e., using both Eq.~\eqref{Eqn:epsilon_jj_IP} for the current-based approach
and Eq.~\eqref{Eqn:epsilon_rr_IP} for the density-based approach.
In doing so we will show potential pitfalls in the implementation of current-based approaches due to:
\emph{(i)} the use of a smearing parameter (Sec.~\ref{sec:smearing}); 
\emph{(ii}) the violation of the conductivity sum rule (Sec.~\ref{sec:drude});
\emph{(iii)} the inclusion of lifetimes (Sec.~\ref{sec:lifetimes});
\emph{(iv)} the inclusion of QP energy corrections and excitonic effects with an incorrect velocity operator (Sec.~\ref{sec:exc}).
Moreover, we demonstrate how those pitfalls can be avoided.

\subsection{Computational details}
%********************************************************************************
The response functions entering Eqs.~\eqref{Eqn:epsilon_jj_IP} and \eqref{Eqn:epsilon_rr_IP} are constructed starting from ground state DFT calculations
performed with either the Quantum-Espresso code\cite{Giannozzi2009} (Si and Cu) or the
Abinit code\cite{abinit} (LiF) using norm conserving pseudopotentials.
The ground-state of all three materials have been calculated by using the local density approximation (LDA) for the exchange-correlation functional. For Si we used 4 valence electrons (${3s^2 3p^2}$ configuration), a face-centered cubic (FCC) cell with a two atoms (diamond structure) and the experimental lattice parameter of $10.18$ Bohr. For calculating the ground state density we use 
an energy cutoff of 10 Ha and a $6\times 6\ \times 6$ sampling of the
Brillouin zone (BZ). For the calculation of the
response function we used instead a $16\times 16\times 16$ sampling of the BZ, resulting in 145 k-points in the irreducible BZ (IBZ) and 4096 in the BZ. For LiF we used 8 valence electrons (${2s^1}$ configuration for Li and ${2s^2 2p^5}$ for F)\cite{Goedecker1996},
a FCC cell with two atoms (sodium-chloride structure) and the experimental lattice parameter of $7.70$ Bohr. For calculating the ground state density and the response function we use an energy cutoff of 40 Ha with a $8\times 8\times 8$ sampling of the BZ, resulting in 29 (512) k-points in the IBZ (BZ). For Cu we used with 11 valence electrons (${3d^{10}4s^1}$ configuration),
a FCC cell with a single atom and the experimental lattice parameter of $6.82$ Bohr.
For calculating the ground state density and the response function we use 32.5 Ha and a $16\times 16\times 16$ sampling of the BZ, resulting in
145 (4096) k-points in the IBZ (BZ).

We computed the dielectric function starting from the DFT-KS wave functions and energies
using the Yambo code~\cite{Marini2009} where we implemented the equations for the current-based approach.\footnote{Yambo standard implementation uses a density-based approach}.
For Si and LiF, we considered electron-hole pairs built from the top 3 valence and the bottom 3 conduction bands. For Cu we included 30 bands in the band summation.  
A scissor operator is used to mimic the effect of the $GW$ quasiparticle corrections in Si
($0.8$ eV) and LiF ($5.8$ eV) consistently to what already reported in the literature. 
For the BSE calculations the static screening in the random-phase approximation is computed
 using 50 bands in the band summation and 2.3 Ha energy cutoff in the summation over reciprocal lattice vectors for Si,
and 30 bands and a 3.6 Ha energy cutoff for LiF.

We also consider QP lifetimes by introducing an imaginary part in the 
QP energies. To mimic the effect of the electron-phonon Fan self-energy~\cite{Giustino2016,Bernardi2014},
we use a term proportional to the density of states. To mimic the effect of the $GW$ self-energy~\cite{Aryasetiawan1998}
we use a term which grows quadratically in energy, from the Fermi level in Cu and from the conduction band maximum
(valence band minimum) plus (minus) the band gap in Si.

\subsection{Smearing parameter and conductivity sum rule} 
\label{sec:smearing}
%********************************************************************************
The macroscopic density-density and paramagnetic current-current response functions 
in Eq.~\eqref{Eqn:chi_0} contain the infinitesimal $\eta$ which ensures causality and avoids
having poles on the real axis.
Numerically $\eta$ can be used as a smearing parameter to simulate finite lifetime effects of the excitations.
As a result each peak in the absorption spectrum acquires a finite width.
This is done by replacing
\be
\w^+ = \lim_{\eta\rightarrow 0}(\w + i\eta) \rightarrow z = \w + i\eta
\label{Eqn:smearing}
\ee
in Eq.~\eqref{Eqn:chi_0} for both the density-based and the current-based approach,
where $\eta$ is now a finite number. In the current-based approach however the frequency $\w$
enters also in the definition of the dielectric function, Eq.~\eqref{Eqn:epsilon_jj_IP}, as a factor
$1/\w^2$. Moreover it is common to numerically impose~\cite{Berger2006,Raimbault2015} the CSR (Eq.~\ref{Eqn:CSR})
replacing the diamagnetic term, $N/V$, with minus the paramagnetic term, $-\x^{\IP}_{j^pj^p}(\w)$, evaluated at $\w=0$.
It is thus natural to wonder whether we should replace also $1/\w^2$ by $1/z^2$ and
if we should use $-\x^{\IP}_{j^pj^p}(z)$ evaluated at $\w=0$ or $z=0$ while imposing the CSR.

\begin{figure}[t]
\centering
\subfigure{\includegraphics[width=8.cm]{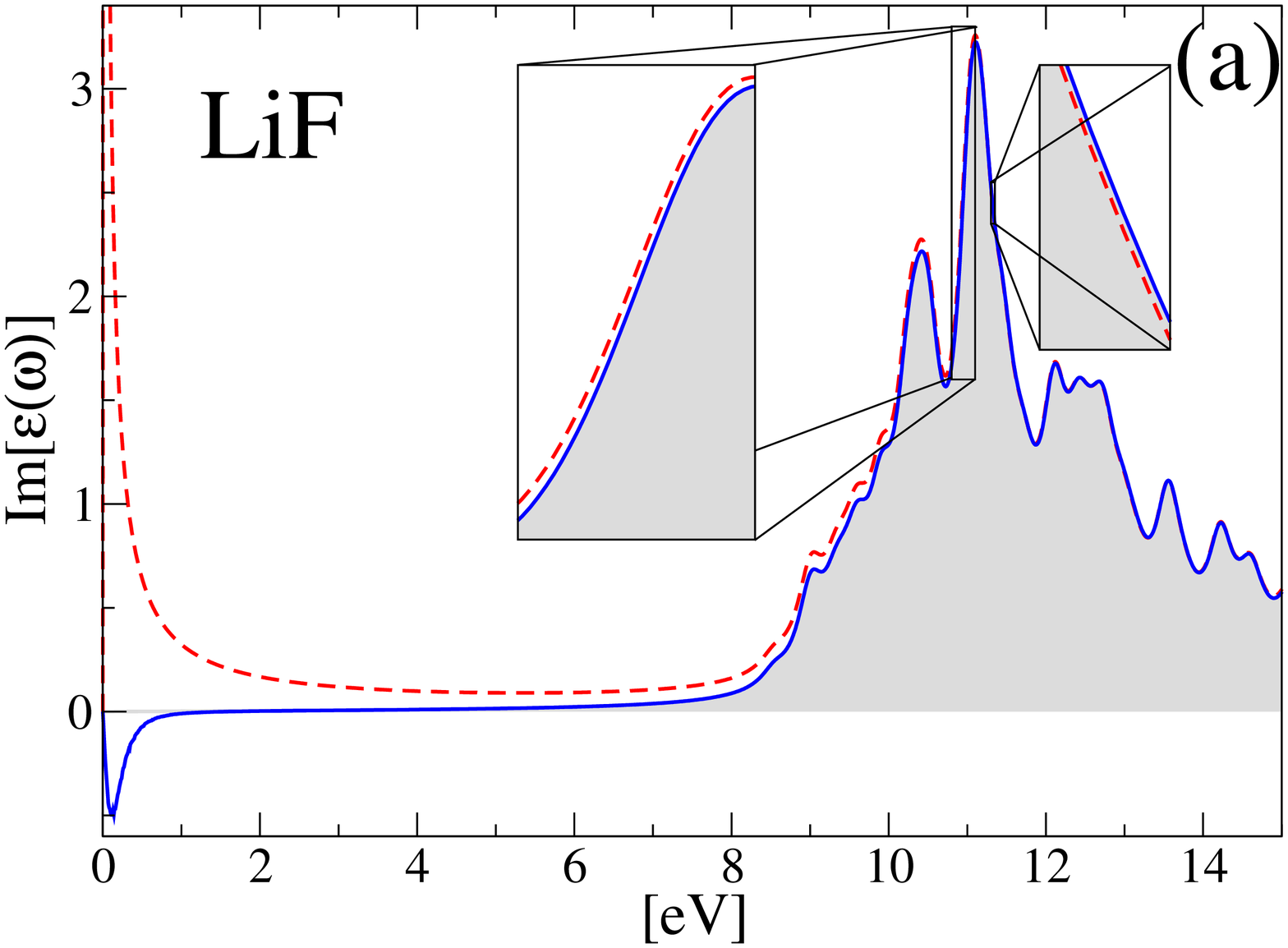}} 
\subfigure{\includegraphics[width=8.cm]{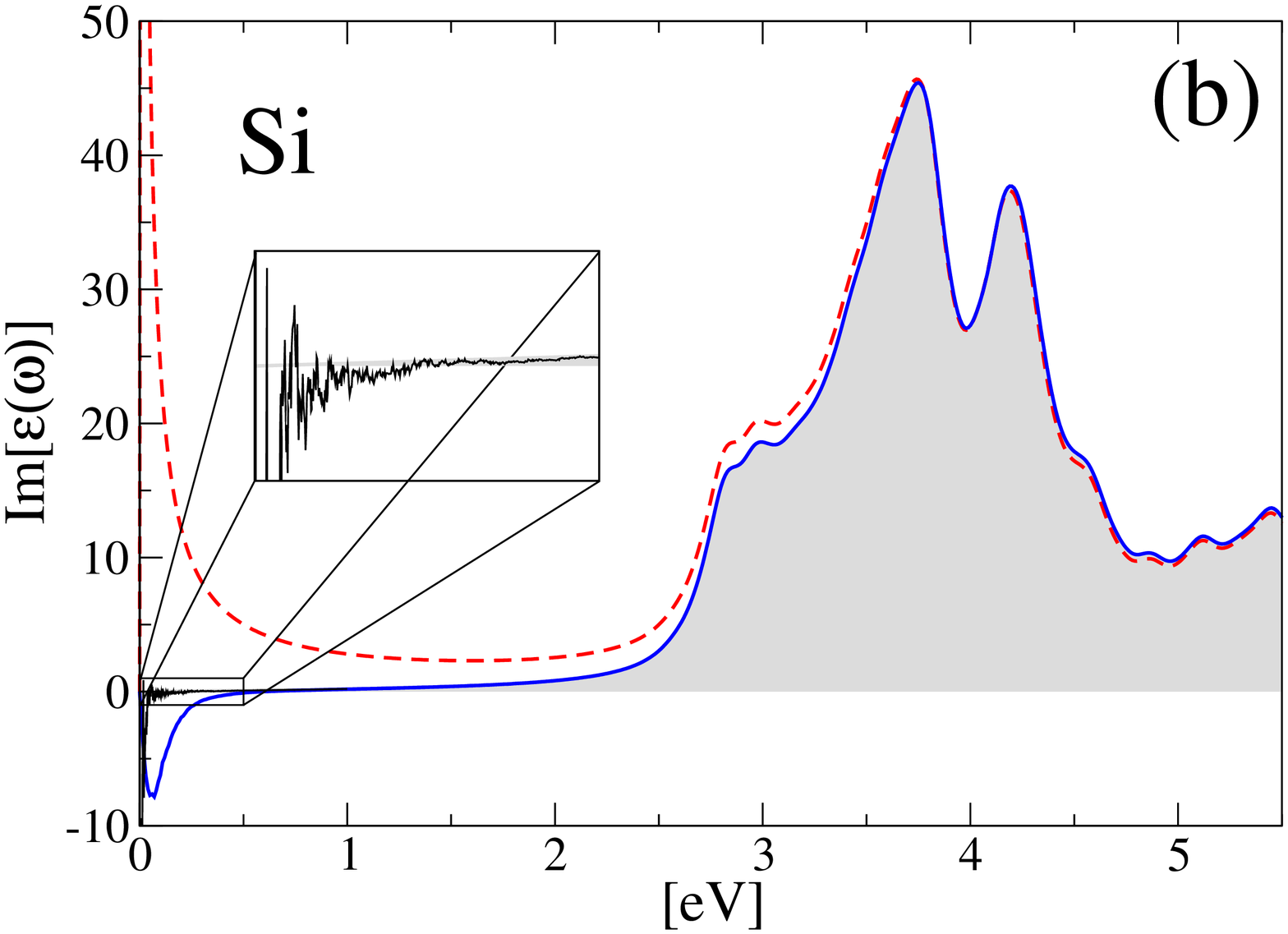}} 
\caption{(color online) Optical absorption in bulk LiF (panel a) and Si (panel b) at the IP level.
   The spectra obtained by replacing $\w^+$ by $\w+i\eta$ in Eq.~\eqref{Eqn:chi_0}, 
   for the density-based approach, Eq.~\eqref{Eqn:epsilon_rr_IP}, (grey shadow)
   and the current-based approach, Eq.~\eqref{Eqn:epsilon_jj_IP} with either $1/\w^2$
   (red dashed line) or $1/z^2$ (blue continuous line), are compared.
   Results with the numerical recipe $\overline{z}=\sqrt{\w+2i\w\eta}$ are also shown.
   Here $\h=0.1$ eV for Si and $\h=0.2$ eV for LiF.
   The conductivity sum rule is always enforced as explained in the main text.}
\label{fig:eta_smearing}
\end{figure}
In Fig.~\ref{fig:eta_smearing} we plot the optical spectra ($\qq=0$) obtained for LiF and Si inserting
$\x^{\IP}_{\rho\rho}(z)$ in Eq.~\eqref{Eqn:epsilon_rr_IP} and $\x^{\IP}_{j^pj^p}(z)$ in
Eq.~\eqref{Eqn:epsilon_jj_IP}. For the current-based approach we consider both the factors
$1/\w^2$ and $1/z^2$ in the definition of the dielectric function (Eq.~\eqref{Eqn:epsilon_jj_IP}).
In both cases the diamagnetic term is replaced by $-\x^{\IP}_{j^pj^p}(i\eta)$, i.e. the CSR is
imposed with $\w=0$.
The optical spectra obtained within the current-based approach are different from those obtained
within the density-based approach. In particular they present clearly unphysical features:
the case with the $1/\w^2$ factor in Eq.~\eqref{Eqn:epsilon_jj_IP} shows a divergent low energy contribution
which resembles the Drude like behaviour of metals; the case with the $1/z^2$ factor shows
a negative peak at $\w=\eta$. Both unphysical features are related to the existence of a finite smearing
in the low frequency region of the spectrum.

One possible solution is to adopt the recipe proposed
by Cazzaniga \textit{et al.}~\cite{Cazzaniga2010} in another context, i.e. to choose
$\tilde{z}=\sqrt{\w^2+2i\h\w}$, instead of $z=\w+i\h$. The latter choice implies no smearing at 
$\w=0$ and a CSR uniquely defined by $N/V=-\x^{\IP}_{j^pj^p}(0)$. For the case of Si we
show this recipe cures both unphysical features although some residual numerical noise remains.

A more rigorous solution requires to
consider again the expansion defined by Eq.\eqref{Eqn:exact_TS}.
Having a smearing parameter we can now expand either around $\w=0$
or around $z=0$ (i.e. $\w=-i\eta)$).
The expansion around $\w=0$ is more suited to analyze the case with the $1/\w^2$ factor in  Eq.~\eqref{Eqn:epsilon_jj_IP} and yields
\begin{equation}
\ve[\x^{\IP}_{jj}](\w)=1+\frac{1}{\w^2}A_{\eta}^{\IP}+\frac{1}{\w}B_{\eta}^{\IP}+C^{\IP}(\w+i\eta).
\end{equation}
The paramagnetic term entering $A_{\eta}^{\IP}$ is correctly balanced by replacing
$N/V\ra-\x^{\IP}_{j^pj^p}(i\eta)$ in the diamagnetic term.
However $B_\eta$ is not zero anymore (it is zero only for $\eta=0$) and thus leads to a divergence in the spectrum at $\w = 0$.

The expansion around $z=0$ is more suited to analyze the case with the $1/z^2$ factor in  Eq.~\eqref{Eqn:epsilon_jj_IP} and yields
\begin{equation}
\ve[\x^{\IP}_{jj}](z)=1+\frac{1}{z^2}A^{\IP}+\frac{1}{z}B^{\IP}+C^{\IP}(z).
\end{equation}
In this case $B^{IP}$ is numerically zero as expected theoretically, however the 
the paramagnetic term entering $A^{\IP}$ is not correctly balanced (since we are using
$N/V\ra-\x^{\IP}_{j^pj^p}(i\eta)$) and the CSR is broken.
$A^{\IP}$ is here multiplied by $1/z^2$ leading to a negative energy peak around $\w=\h$.
To summarize, in order to avoid unphysical divergencies and negative peaks there are two options:
\begin{itemize}
\item the $1/\w^2$ factor can be used in Eq.~\eqref{Eqn:epsilon_jj_IP} together with the CSR imposed by $N/V\ra-\x^{\IP}_{j^pj^p}(i\h)$
and balancing $B_{\eta}^{\IP}$ by a proper counter term;
\item a $1/z^2$ term can be used in Eq.~\eqref{Eqn:epsilon_jj_IP} together with the CSR imposed by $N/V\ra-\x^{\IP}_{j^pj^p}(0)$.
\end{itemize}
The latter option is the most straightforward to implement and we have tested that cures the unphysical negative peak at $\omega = \h$ and gives the same spectra as within the density-based approach.

\subsection{The conductivity sum rule and the Drude term} 
\label{sec:drude}
%********************************************************************************
The current-current response function in Eq.~\eqref{Eqn:CCD} consists of two terms,
a constant diamagnetic term $N/V$ and a paramagnetic term given by Eq.~(\ref{Eqn:chi_0}).
In practice the sum over states in Eq.~(\ref{Eqn:chi_0}) is truncated.
As a consequence the CSR in Eq.~(\ref{Eqn:CSR}) is no longer satisfied,
and it becomes impossible to converge the optical spectra for small frequencies.
To solve these problems one can impose the CSR by
replacing~\cite{Kootstra2000,Romaniello2005,Berger2005}
the diamagnetic term $N/V$ with $-\x_{j^pj^p}(\qq,0)$
(or $-\x_{j^pj^p}(\qq,i\eta)$ as we did in the previous section).
Thus the diamagnetic and paramagnetic contributions are treated on
equal footing and no convergence problems occur.

\begin{figure}[t]
\centering
\subfigure{\includegraphics[width=8.cm,height=6.25cm]{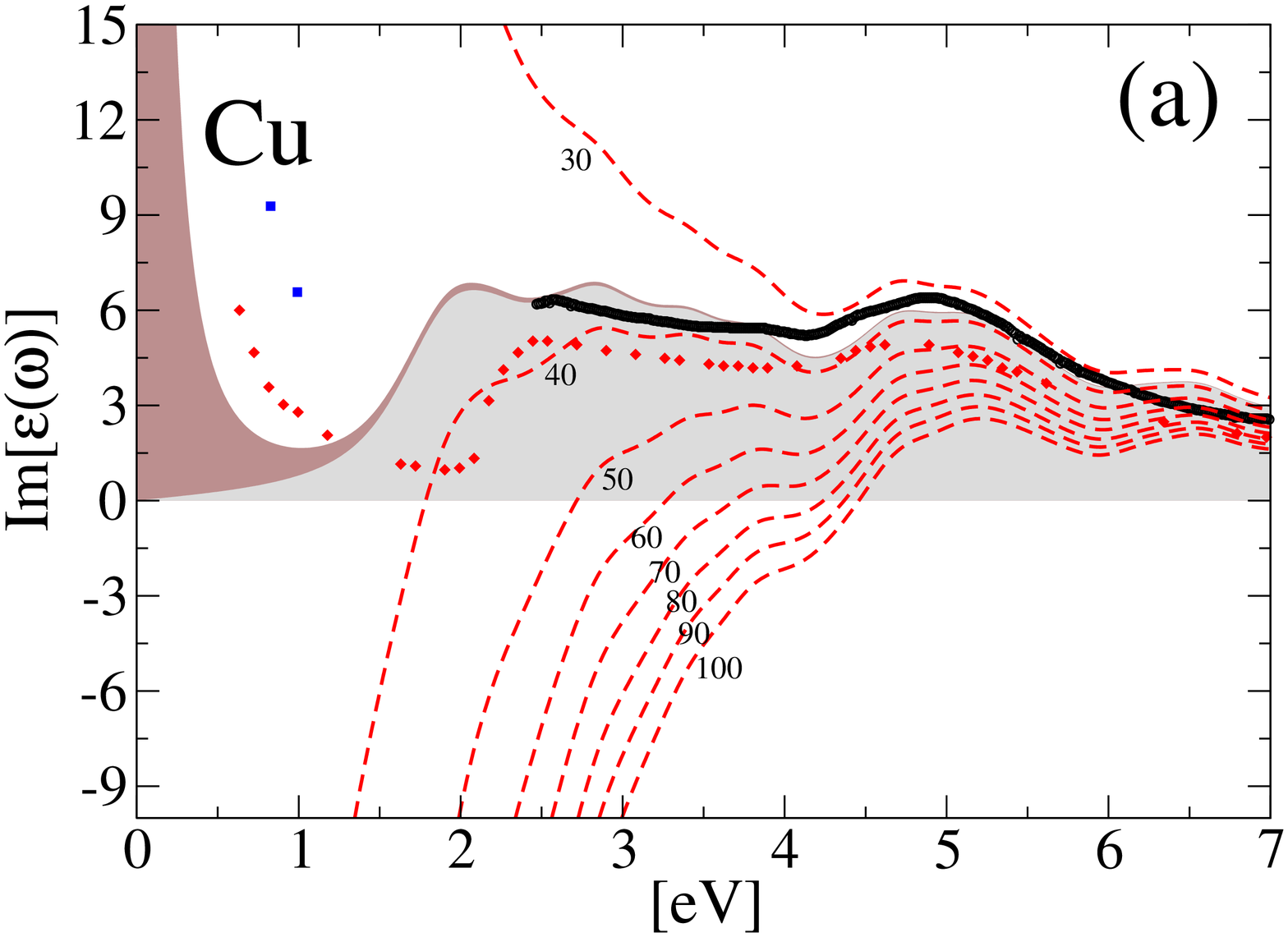}}
\subfigure{\includegraphics[width=8.cm,height=6.25cm]{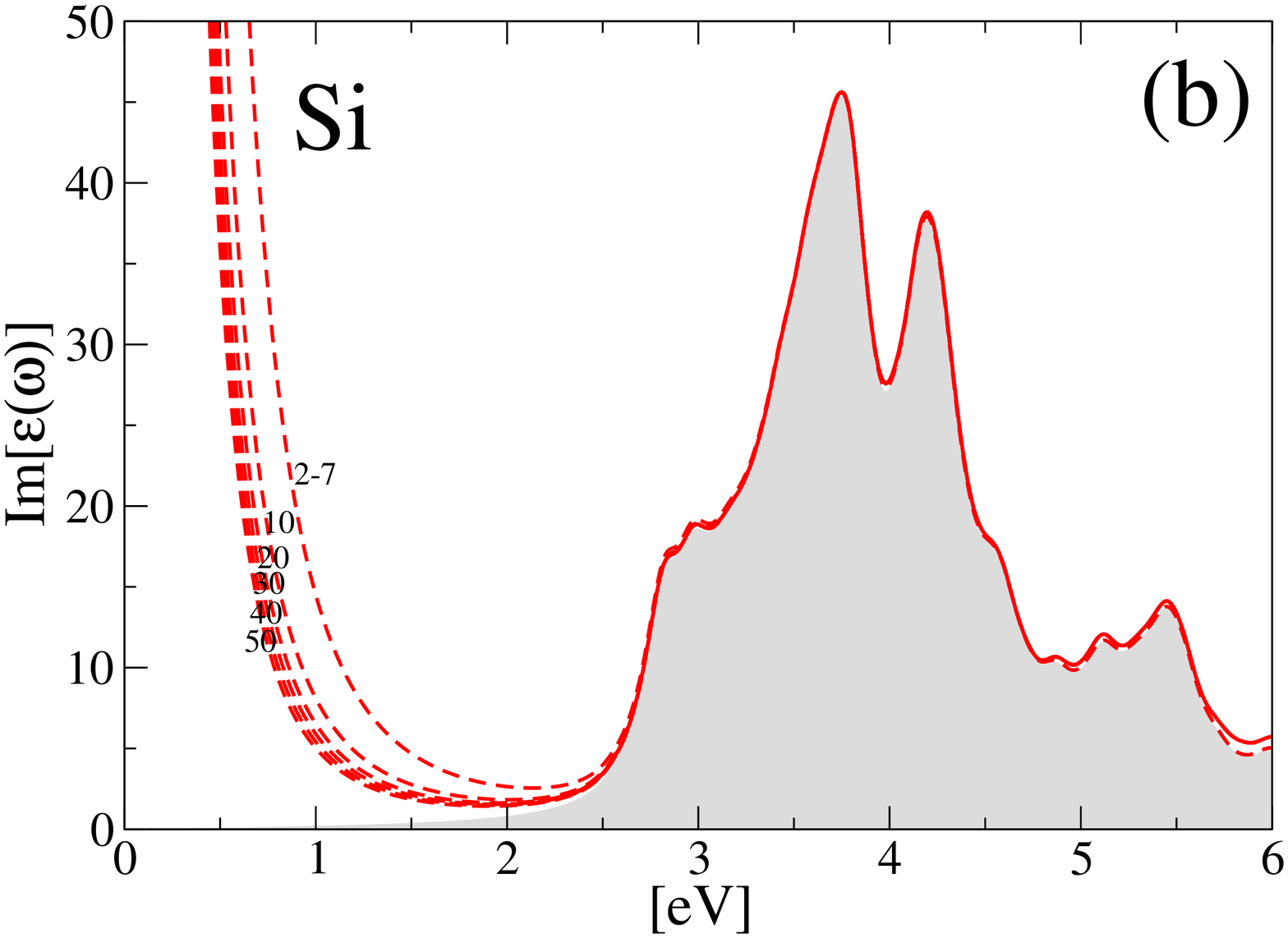}}
\caption{(color online) Optical absorption in bulk Cu (panel a) and Si (panel b) at 
  the IP level. Red dashed lines: spectra obtained in  the current-based approach using Eq~\eqref{Eqn:current_full} without enforcing the conductivity sum rule (Eq~\eqref{Eqn:CCD}); the various lines correspond to different number of bands. Grey shadow: spectra obtained in the density-based approach (Eq.~\eqref{Eqn:epsilon_rr_IP}) and in the current-based approach by enforcing the conductivity sum rule (Eq~\eqref{Eqn:CCD}). Brown shadow: Drude term added via a Drude model (for Copper). Blue dots, red dots, and black continuous bold line: experimental data from Ref.~\onlinecite{Dold1975}, Ref.~\onlinecite{ehrenreich}, and Ref.~\onlinecite{stahrenberg}, respectively.}
  \label{fig:CSR}
\end{figure}

At $\qq=\zero$ this strategy poses no problems for systems with a gap. 
For metals, instead, it suppresses the Drude tail, which is described by the term $A^{IP}/\w^2$ on the right-hand side of Eq.\ (\ref{Eqn:current_full}).
This occurs because the CSR in general does not hold at $\qq=\zero$ since intraband transitions are excluded in the sum over states in $\chi_{j^pj^p}$.
We thus consider here the case of a metal, Cu, and compute its spectrum without imposing the CSR.
Since the diamagnetic term is purely real, we need to use the $1/z^2$ strategy as discussed previously.
However, without imposing the CSR, calculations never converge. 
This is shown in Fig.~\ref{fig:CSR}: the Drude tail in Cu has a wrong behaviour, whereas in Si, where converged calculations 
should give no absorption at low energy, an artificial Drude-like tail appears. 

We found that the best solution is to enforce the 
CSR and to calculate the Drude
term through the explicit inclusion of intraband transitions as described for the density-based approach. 
A faster convergence of the latter contribution can be obtained using the
tetrahedron method for the integration in the Brillouin zone.~\cite{Blochl1994,Romaniello2005}

\subsection{Many-body lifetimes}
\label{sec:lifetimes}
%********************************************************************************

In Sec.~\ref{sec:smearing} the finite width of the peaks was obtained
by introducing an \emph{ad hoc} smearing parameter.
A more physically motivated approach is to consider finite
lifetimes $\g$ originating from the imaginary parts of the many-body self-energy.
For example, it has been shown that in a semiconductor such as Si the finite width of the peaks
is well-described by means of the Fan and $GW$ self-energies, which account for electron-phonon\cite{Giustino2016,Marini2008}
and electron-electron~\cite{Aryasetiawan1998} scattering  processes, respectively.
To illustrate the effects of such lifetimes in the following we will consider the contributions
due to the Fan self-energy. We define modified KS energies $\et^{\KS}_{n}$ as
\be
\et^{\KS}_{n}=\e^{\KS}_{n}+i\g_{n},
\label{Eqn:complex_enenrgies}
\ee
and use them to compute the absorption both in the current-based
and in the density-based approach. 
The KS energies $\et^{\KS}_{n}$ correspond to the
eigenenergies of the Hamiltonian (\ref{Eqn:QPH}), in which only the imaginary part of the Fan
self-energy is considered. The underlying Hamiltonian, which we will indicate as $\tilde{H}_{\KS}$,
is diagonal in the same basis set in which $H_{\KS}$ is diagonal,
i.e. $\tilde{\psi}^{\KS}_{n}\approx\psi^{\KS}_{n}$ and thus $\tilde{x}^{\KS}_{nm}=x^{\KS}_{nm}$.
The velocity matrix elements however change because the velocity operator is proportional to
the hamiltonian itself. 
Using Eq.~\eqref{Eqn:velocity_operator_general}, the velocity matrix elements read
\bea
\tilde{v}^{\KS}_{nm}&=& -i\, \tilde{x}^{\KS}_{nm}\ [\et^{\KS}_{n}-(\et^{\KS}_{m})^*]\notag  \\
           &=& v^{\KS}_{nm} \frac{\et^{\KS}_{n}-(\et^{\KS}_{m})^*}{\e^{\KS}_{n}-\e^{\KS}_{m}}.
\label{Eqn:dressed_KS_velocity} 
\eea
Such expression generalizes the result by Del Sole et al.\cite{DelSole1993} (that we use in the next section) to the case of complex energies and is consistent with the findings of Tokman.~\footnote{Ref.~\onlinecite{Tokman2009} discusses the connection between the time derivative of the dipole
operator and the velocity operator in case of a dephasing of the
polarization (i.e. an imaginary term in the Hamiltonian) is considered.}

\begin{figure}[t]
\centering
\subfigure{\includegraphics[width=8.cm,height=6.25cm]{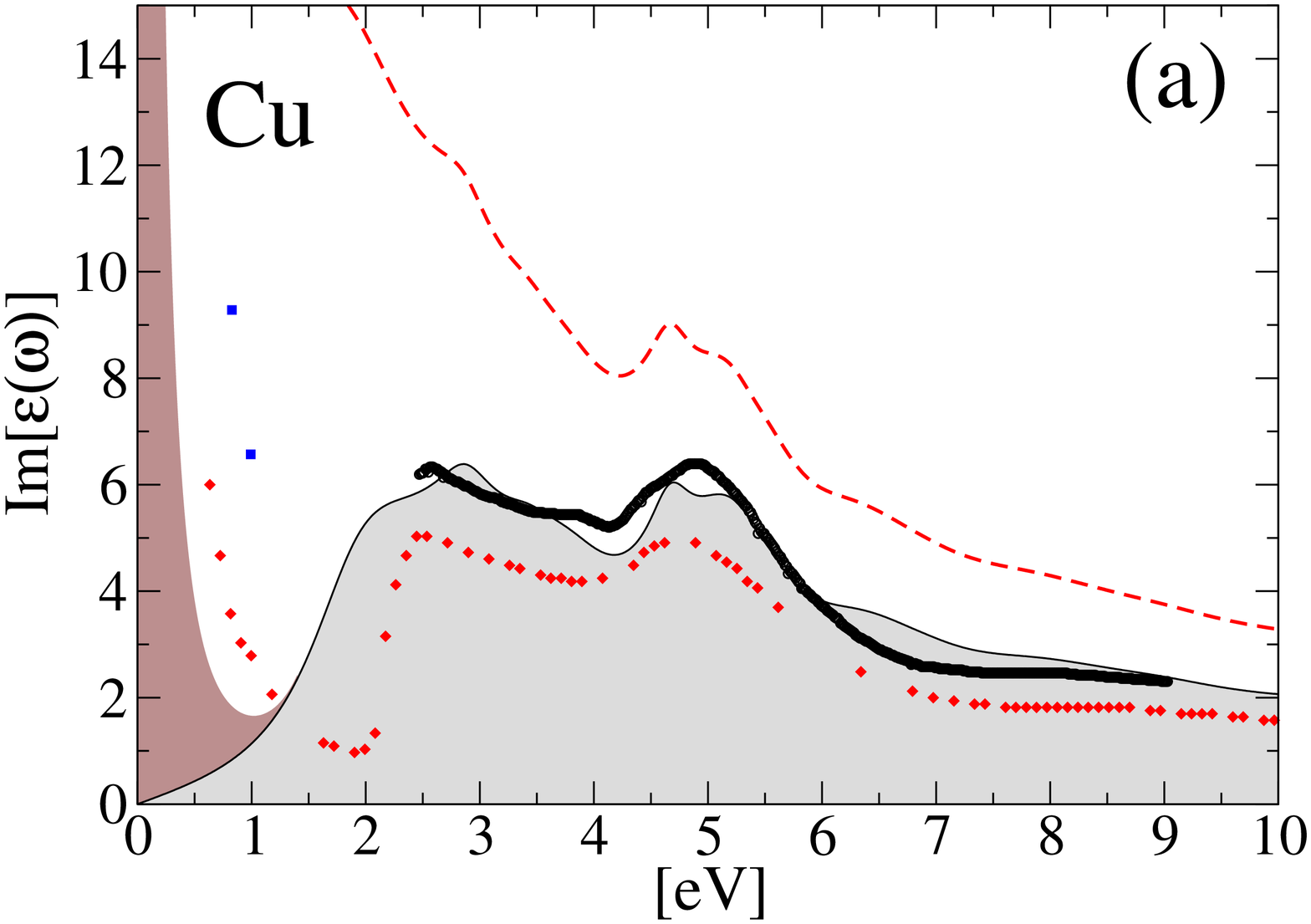}}
\subfigure{\includegraphics[width=8.cm,height=6.25cm]{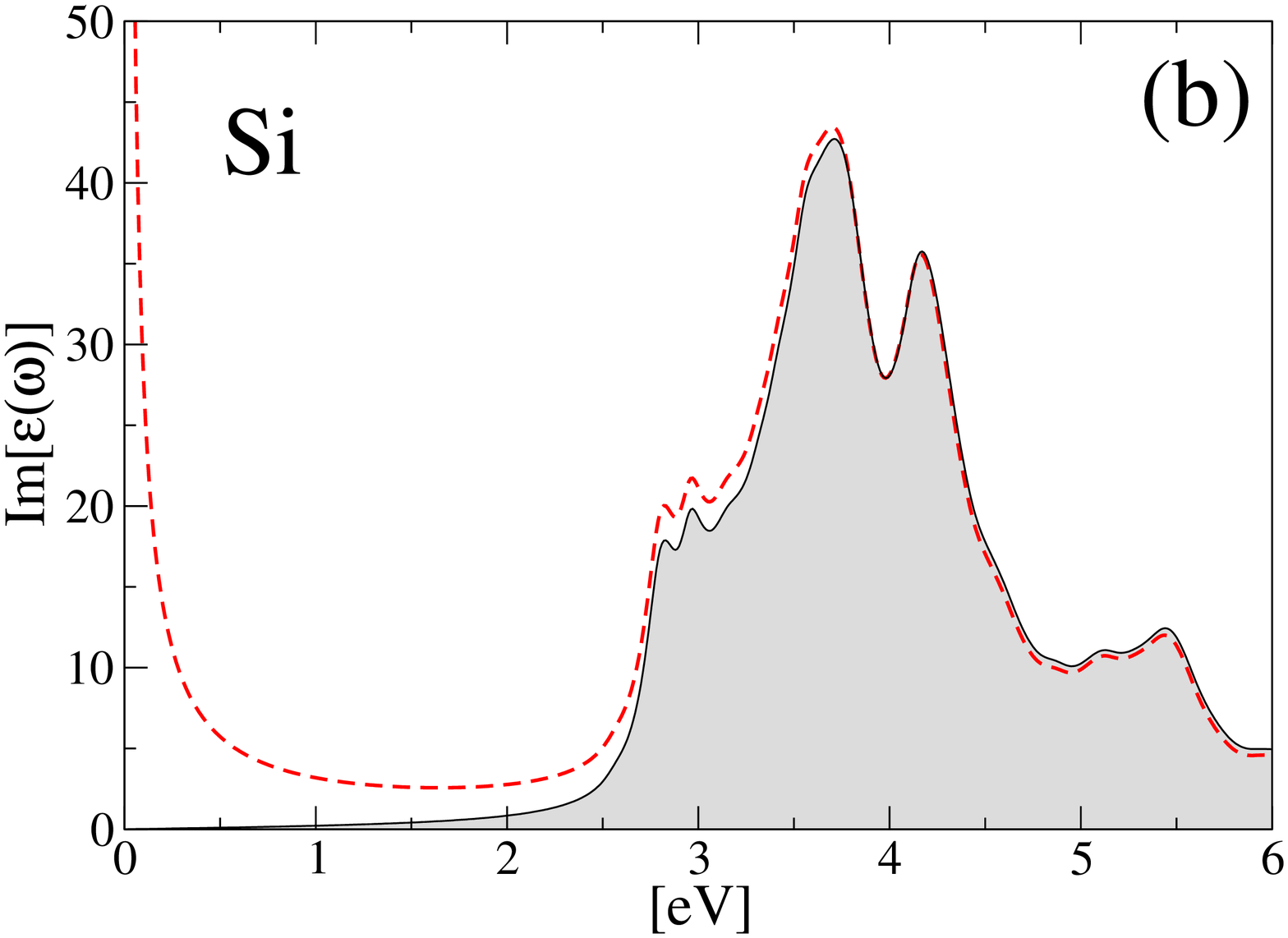}}
\caption{(color online) Optical absorption in bulk Cu (panel a) and Si (panel b) at the IP level. 
Red dashed line: spectra obtained in the current-based approach using the complex one-particle energies (\ref{Eqn:complex_enenrgies}) in Eq.~\eqref{Eqn:chi_0}. Grey shadow: spectra obtained in the density-based approach (Eq.~\eqref{Eqn:epsilon_rr_IP}) and in the current-based approach by using the complex one-particle energies (\ref{Eqn:complex_enenrgies}) and the velocity operator \eqref{Eqn:dressed_KS_velocity} (with \eqref{Eqn:v_non_herm}) in Eq.~\eqref{Eqn:chi_0}. 
Brown shadow: as in Fig.~\ref{fig:CSR}
Blue dots, red dots, and bold continuous black line: as in Fig.~\ref{fig:CSR}.} 
\label{fig:QP_lifetimes}
\end{figure} 

Moreover, since $\tilde{H}_{\KS}$ is non-Hermitian (although it is still a normal matrix), the
velocity operator is also non-Hermitian. This means in practice
\be
\tilde{v}^{\KS}_{nm} \neq \(\tilde{v}^{\KS}_{mn}\)^*,
\label{Eqn:v_non_herm}
\ee
and thus the numerator of the response function (Eq.~\eqref{Eqn:chi_0}) in the current-based approach cannot be 
written as a square modulus $|\tilde{v}^{\KS}_{nm}|^2$ anymore. 
One may wonder if---having a non positive defined numerator---the spectrum may become negative.

In Fig.~\ref{fig:QP_lifetimes} we show for Cu and Si that this is not the case: the spectrum is well defined and matches the one obtained in the density-based approach as long as the non-Hermiticity of the velocity operator is correctly taken into account.
If instead we ignore the non-Hermicity of the velocity operator and use  $|\tilde{v}^{\KS}_{nm}|^2$ in the numerator of Eq.~\eqref{Eqn:chi_0}, a spurious divergence appears at low energy for Si and the Drude tail is not correctly described for Cu.

\subsection{Quasiparticle energies and excitonic effects\label{Sec:EXC}}
\label{sec:exc}
%********************************************************************************

We finally consider the dielectric function beyond the IP approximation by using the Bethe-Salpeter equation. We include both the $GW$ corrections to the KS band structure and
the effect of the electron-hole interaction in the absorption. 
As for the case of the QP lifetimes we have to consider a renormalized 
velocity operator,~\cite{DelSole1993}
\bea
v^{\QP}_{nm}&=& -i x^{\QP}_{nm}\ (\e^{\QP}_{n}-\e^{\QP}_{m}) \\
           &=& v^{\KS}_{nm} \frac{\e^{\QP}_{n}-\e^{\QP}_{m}}{\e^{\KS}_{n}-\e^{\KS}_{m}},
\label{Eqn:QP_velocity} 
\eea
which corresponds to the quasiparticle Hamiltonian (Eq.\eqref{Eqn:QPH}) with a $GW$ self-energy.
However this is not the only correction to be considered.
If the dielectric function is computed by
diagonalizing the excitonic Hamiltonian given by Eq.~(\ref{Eqn:H2p_eigen}), then it is expressed
in terms of the excitonic dipole matrix elements $x^\Exc_\l$ in the density-based approach,
\be
x^{\Exc}_\l=\sum_{nm} A^\l_{nm} x^{\QP}_{nm},
\label{Eqn:exc_dip_mat}
\ee
or via the excitonic velocity matrix elements $v^\Exc_\l$, in the current-based approach.
One might be tempted, in analogy with Eq.~(\ref{Eqn:exc_dip_mat}), to define 
\be
v^{\Exc}_\l=\sum_{nm} A^\l_{nm} v^{\QP}_{nm}. 
\label{Eqn:BSE_velocity_wrong}
\ee
However this expression---which has been reported in the literature \cite{Louie2000,Spataru2005,Palummo2015}---is not correct 
since it does not correspond to the underlying excitonic Hamiltonian $H^{\Exc}$ given by Eq.~(\ref{Eqn:H2p_eigen}). 
The excitonic velocity hence must read
\bea
v^{\Exc}_\l&=&-i\,\bra [H^{\Exc},x] \ket=-i\,x^{\Exc}_\l\ E_\l \label{Eqn:BSE_commutator}\\
          &=&\sum_{nm} A^\l_{nm} v^{\QP}_{nm} \frac{E_\l}{\e^{\QP}_{n}-\e^{\QP}_{m}}
\label{Eqn:BSE_velocity}
,
\eea
which in general is different from Eq.~(\ref{Eqn:BSE_velocity_wrong}).

\begin{figure}[t]
\centering
\subfigure{\includegraphics[width=8.cm]{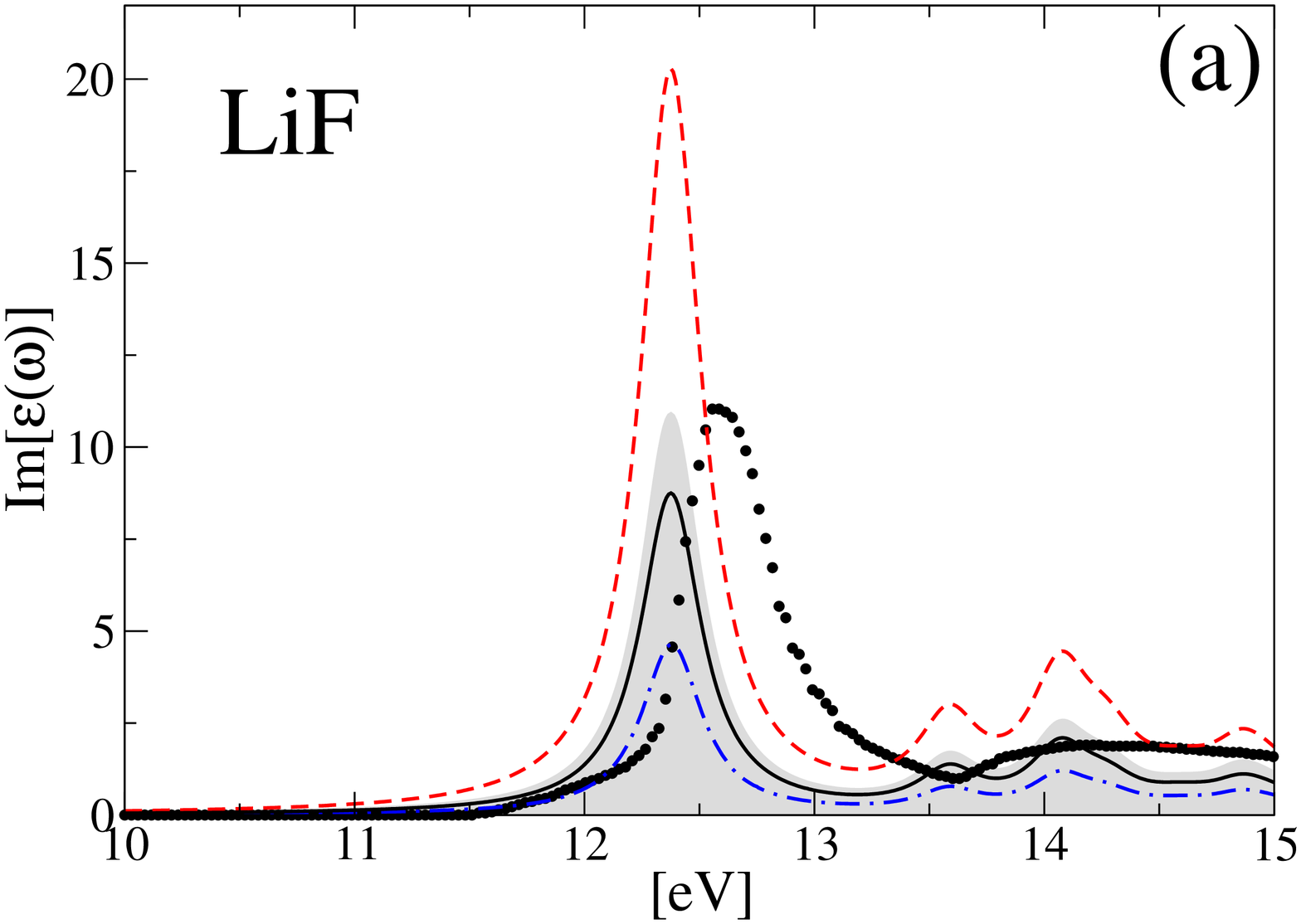}}
\subfigure{\includegraphics[width=8.cm]{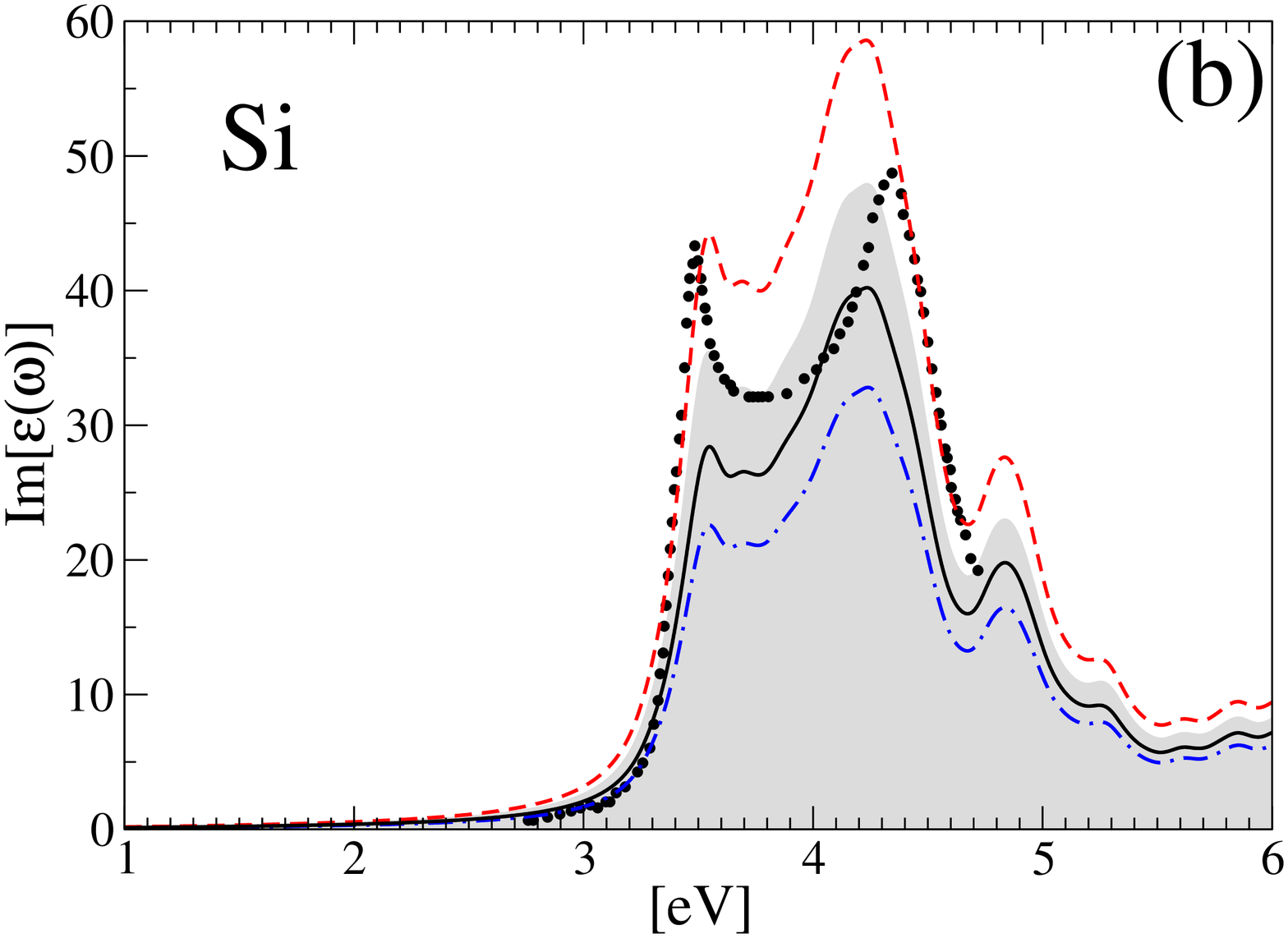}}
\caption{(color online) Optical absorption in bulk LiF (panel a) and Si (panel b) at 
the BSE level. Black continuous line: spectra obtained in the current-based approach (Eq.~\eqref{Eqn:epsilon_jj_IP}) neglecting the corrections \eqref{Eqn:QP_velocity} and \eqref{Eqn:BSE_velocity}.  Red dashed line: spectra obtained in the current-based approach (Eq.~\eqref{Eqn:epsilon_jj_IP}) considering only the correction \eqref{Eqn:QP_velocity}. Blue dot-dashed line: spectra obtained in the current-based approach (Eq.~\eqref{Eqn:epsilon_jj_IP}) considering only the correction \eqref{Eqn:BSE_velocity}. Grey shadow: spectra obtained in the density-based approach (Eq.~\eqref{Eqn:epsilon_rr_IP}) and in the current-based approach considering the corrections \eqref{Eqn:QP_velocity} and \eqref{Eqn:BSE_velocity}. Black dots: experimental data are from Ref.~\onlinecite{lautenschlager1987temperature,roessler1967optical}.}
\label{fig:Excitonic_effect}
\end{figure}

In Fig.~\ref{fig:Excitonic_effect} we compare the absorption spectrum in the density-based
approach with the current-based approach results with and without the renormalized
velocity operator, i.e. considering the corrections due to Eqs.\eqref{Eqn:QP_velocity}
and \eqref{Eqn:BSE_velocity}.
We notice that the two corrections in general partially cancel each other since
$\e^{\QP}_{n}-\e^{\QP}_{m}>\e^{\KS}_{n}-\e^{\KS}_{m}$ and $E_\l<\e^{\QP}_{n}-\e^{\QP}_{m}$
and the resulting error in the absorption intensity is proportional to $|E_\l - (\e^{\KS}_{n}-\e^{\KS}_{m})|/|\e^{\KS}_{n}-\e^{\KS}_{m}|$.
There is thus an error cancellation in the intensity due
to the fact that the onset of the KS absorption is often not too different
from the onset of the BSE absorption.
The opposite sign of the two corrections is evident when they are considered independently.
If only Eq.~\eqref{Eqn:QP_velocity} is considered, the absorption spectrum intensity
is strongly overestimated. Conversely the spectrum intensity is strongly underestimated
if only Eq.~\eqref{Eqn:BSE_velocity} is considered.
Only when including both the renormalization of the velocity due the quasiparticle corrections and the nonlocal operator we recover the results obtained within the density-based approach.

Regarding the computational cost, we note that if Eq.~(\ref{Eqn:BSE_velocity}) is used to evaluate the excitonic velocity operator $v^{\Exc}_\l$, then
iterative schemes to calculate the dielectric function, such as the Lanczos-Haydock method~\cite{Haydock1980,Benedict1999,Gruning2009}, are precluded since $v^{\Exc}_\l$ in Eq.~(\ref{Eqn:BSE_velocity}) explicitly depends on the BSE energies $E_\l$.
Instead, one should use Eq.~(\ref{Eqn:BSE_commutator}) and explicitly calculate the commutator of the dipole operator and the excitonic Hamiltonian $H^{\Exc}$ which is cumbersome since $H^{\Exc}$ contains nonlocal kernels.
This issue also arises in approaches in which the QP corrections are calculated
in a non-perturbative way, i.e., when a non-diagonal self-energy is considered, and thus
the commutator with such a self-energy should in principle be computed.~\cite{Bruneval2006}

\section{Direct calculation of the induced density, polarization and current\label{sec:Induced_quant}}
%********************************************************************************

For sake of completeness we extend the discussion by considering alternative approaches in which optical response functions are not
computed as sum over states (Eq.~\ref{Eqn:chi_0}), but rather using the right-hand
side of Eqs.(\ref{Eqn:chi_IP}): the change
in the density or in the current-density induced by a perturbing field~\cite{Bertsch2000,Gruning2016,Yabana2012} is computed first and 
the macroscopic response functions are then obtained by dividing the induced
quantities by the perturbing fields.

A way to access the density and the current-density is through the one particle
time-dependent density-matrix of the system $\varrho(\blx,\blx',t)$. 

The general equation of motion (EOM) for the density matrix can be written
in the Von Neumann form~\cite{Neumann1927}. It reads
\be
\hat L(t)\, \g(t) - \hR(t)\, \varrho(t) =0  \label{vonneq},
\ee
where the Liouvillian operator $\hat L(t)$, defined by
\be
\hat L(t) \varrho(t) =i\partial_t\,\varrho(t) - [ h_+(t),\varrho(t) ], \label{coherent}
\ee
describes the coherent evolution in terms of the Hermitian part of the Hamiltonian $\hat{h}_+(t)$.
The relaxation operator $\hat R$, defined by 
\be
\hR(t)\, \varrho(t) = \{h_-(t),\varrho(t)\} \label{relaxsimple},
\ee
describes relaxation processes in terms of the skew-Hermitian part of the Hamiltonian $\hat{h}_-(t)$.\footnote{This EOM can also be derived as an approximation to the non-equilibrium Green's function theory or quantum kinetic theory~\cite{kadanoff,Bonitz} where the density matrix is the time diagonal of the lesser Green function $\varrho(t)=-iG^<(t,t)$.}
We can solve the equation for $\varrho(t)$ in the equilibrium one-particle basis set
$\varrho_{nm}(t)=\langle\psi_m|\varrho(t)|\psi_n\rangle$ with the initial condition $\varrho_{nm}(t=0)=\delta_{nm}f_n$.
The density and current-density can then be obtained as 
\begin{subequations}
\label{Eqn:rho-j_def_neq}
\begin{align}
\r(\blx,t)  &= \sum_{nm} \varrho_{nm}(t)\, \psi^*_n(\blx)\, \hat{\r}\, \psi_m(\blx),      \label{Eqn:rho_def_neq}  \\
\blj(\blx,t)&= \sum_{nm} \varrho_{nm}(t)\, \psi^*_n(\blx)\, \hat{\blv}(t)\, \psi_m(\blx). \label{Eqn:j_def_neq}
\end{align}
\end{subequations}
If the Hamiltonian is expressed in terms of quasiparticle energies
and only the variation of the static screened exchange self-energy is considered, this approach has been proven~\cite{Attaccalite2011}, to linear order, to be equivalent to the BSE. In this case the use of a smearing parameter $\h$ in the BSE corresponds to setting the non-Hermitian part of the Hamiltonian proportional to $\h$.

Since the Hamiltonian is gauge dependent, so is the EOM in Eq.~(\ref{vonneq}) and its solution $\varrho(t)$.
Using the gauge function $\Lambda(\blx,t)$ defined in App.~\ref{App:gauges}, which transforms
the potentials from one gauge to another, $\hat L(t)$, $\hR(t)$, and $\varrho(t)$ transform 
according to 
\bea
\hat L_2 &=& e^{i\chi} \hat L_1  e^{-i\chi} \label{Eqn:gauge_transf_L}  \\
   \hR_2 &=& e^{i\chi}    \hR_1  e^{-i\chi} \label{Eqn:gauge_transf_R}  \\
  \varrho_2 &=& \varrho_1 e^{i \chi}              \label{Eqn:gauge_transf_rho},
\eea
where $\chi = c\,[\Lambda(\blx,t)-\Lambda(\blx',t)]$.
One can show that a gauge transformation preserves the current, the density,
the polarization and the electromagnetic energy~\cite{Tokman2009}
only if all quantities are gauge transformed together. We note that in the velocity gauge
the relaxation operator acquires a dependence on the perturbing potential, not present
in the length gauge.\cite{Tokman2009,Lamb1987}.
At finite momentum Eqs.~\eqref{Eqn:gauge_transf_L}-\eqref{Eqn:gauge_transf_rho} 
are all is needed to prove the equivalence between the density- and current-based approaches for the description of optical
properties, similarly to Eq.~\eqref{Eqn:Xjj_Xrr_relation} for the response functions.
Propagating the density-matrix in the length gauge (with potential $\phi(t)$) and then computing the density variation
the macroscopic density-density response function is obtained.
Propagating the density-matrix in the velocity gauge (with potential $\blA(t)$) and then computing the current-density variation
the macroscopic current-current response function is obtained.

Again difficulties arise if one considers the optical limit $\qq\ra\zero$.
Indeed at $\qq=\zero$ the length gauge cannot be formulated in terms
of the coupling with the density only~\cite{Gonze1995,Martin1997}. The induced macroscopic polarization
needs to be defined.
Up to first order in the perturbing field, it can be constructed~\footnote{for a discussion and a general expression, valid at all orders, see  Ref.~\onlinecite{Attaccalite2013}}
from the density matrix as
\be
P(t)=\sum_{n\neq m} \varrho_{nm}(t) x_{\zero,nm}.\label{linearp}
\ee
The longitudinal~\footnote{While the density-density response function can be used only to compute
the longitudinal response, the dipole-dipole response function, like the current-current one, can also describe
transverse and mixed (longitudinal-transverse) terms.}
dipole-dipole response function can then be defined as
${\x_{dd}(\w)=\d P(\w)/\d E(\w)}$, where $E$ is the total macroscopic electric field.
Thanks to Eq.~\eqref{linearp} it is possible to show the formal equivalence~\cite{Tokman2009}
between the two formalisms also at $\qq=0$ for cold semiconductors.
As previously, $\x_{dd}(\w)$ describes however only the inter-band and not the
intra-band contribution. The latter can be obtained, in the length gauge, only by explicitly
considering the $\qq\ra\zero$ limit. 
The direct numerical calculation of $\d\r(\qq,\w)$ at very small $\qq$ is
however not feasible, and one would need to analytically deal with the small $\qq$ dependence
as we do in App.~\ref{App:intra} for the response function.
Instead, the current-based formalism can be directly formulated also at $\blq=\zero$ and
the intraband contribution is also included. However, similar to what discussed previously, one must be careful because of potential breaking
of the CSR (Eq.~\ref{Eqn:CSR}).

%********************************************************************************
\section{Conclusions\label{Sec:Conclusions}}
%********************************************************************************
We compared the optical absorption of extended systems calculated from the density-density and current-current linear response functions obtained within many-body perturbation theory. 
We did this by studying the longitudinal macroscopic dielectric function both on a formal and on a numerical level.

We showed that for a finite momentum $\qq$, carried by the external perturbation, the two approaches are formally equivalent thanks to the continuity equation, which relates the density to the longitudinal current.

For the case of vanishing momentum, $\qq=\zero$, the optical absorption is not well defined in terms of the density-density response function.  A small $\qq$ expansion is needed, which leads to a formulation in terms of the dipole-dipole response function plus a divergent term which describes intraband transitions for metalic systems. The current-based approach is instead exact at $\qq=\zero$. In practice,  however, one needs to impose the conductivity sum rule in a way which suppresses the intraband transitions at $\qq=\zero$, thus making the small $\qq$ expansion needed for metalic systems also in the current based approach.  

When gapped systems are considered there are no intraband transitions and the two approaches are formally equivalent once the conductivity sum rule is imposed in the current-based approach. On the other hand we showed that the formal equivalence of the approaches may be lost in practical calculations when the strategies usually employed in the density-based approach to include smearing, quasiparticle lifetimes and the electron-hole interaction are naively applied to the current-based approach. The smearing is included straightforwardly in the density-based approach by replacing the real frequency with an imaginary frequency. However, a careless extension of this recipe within the current-based approach leads to unphysical features in the optical spectrum. We showed how correctly include the smearing by redefining the dielectric function and the conductivity sum rule.  The inclusion of  lifetimes, quasiparticle corrections, and excitonic effects correspond to a change in the underlying Hamiltonian. Therefore the velocity operator, which enters the definition of the current-current response function,
has to be modified accordingly. We noted instead that the expression for the excitonic velocity operator reported in several published works is incorrect. In this work we thus report the correct general definition for the velocity operator when complex energies, quasiparticle corrections, and excitonic effects are taken into account.

%********************************************************************************
\section*{Acknowledgments}
%********************************************************************************
The authors thank the EUspec COST Action for Short Term Scientific Missions that allowed to initiate and carry out this project. Discussion within
the Collaboration Team on Correlation of the European Theoretical
Spectroscopy Facility (ETSF) is greatly acknowledged.
DS acknowledge Giovanni Onida for the computational time provided
on the HPC cluster ``ETSFMI'' in Milano and Paolo Salvestrini for the support on the cluster;
computing time has been also provided by the French national GENGI-IDRIS supercomputing centers at Orsay under contract $n^o$ t2012096655;
financial support by the {\em Futuro in Ricerca} grant No. RBFR12SW0J of the
Italian Ministry of Education, University and Research MIUR;
the funding received from the European Union project
MaX {\em Materials design at the eXascale} H2020-EINFRA-2015-1, Grant agreement n. 676598 and
{\em Nanoscience Foundries and Fine Analysis - Europe} H2020-INFRAIA-2014-2015,
Grant agreement n. 654360. JAB and PR acknowledge support from the IDEX Emergence (project no. 2016-075/CNRS)

\section*{Appendices}
%********************************************************************************
\appendix
%********************************************************************************

\section{Fourier Transform in Periodic Boundary Conditions} \label{App:FT_in_PBC}
%********************************************************************************
Here we explicitly introduce the Fourier transform in
Periodic Boundary Conditions (PBC).
To make the derivation more clear, we use, just for the present appendix, different symbols for the
different functions in real and reciprocal space associated to a given observable $O$.
We start from the general definition:
\begin{subequations}
\begin{align}
O(\blx)&=\int \frac{d^3\blk}{(2\pi)^3}\, \mathcal{O}(\blk) e^{i\blk\cdot\blx}, \\
\mathcal{O}(\blk)&=\int d^3\blx O(\blx) e^{-i\blk\cdot\blx}.
\end{align}
\end{subequations}
We then divide the space in a series of microscopic unit cells with volume $V$ which are
periodically repeated and we use $\blx=\blr+\blR$ with $\blr$ restricted to the volume $V$
centered at $\blR=\zero$. Accordingly the reciprocal space results separated in two
parts with $\blG$ defined by $\blG\cdot\blR=2\pi\,n$, where $n$ is any integer number, 
and $\blq=\blk-\blG$ restricted to the first Brillouin Zone
with volume ${\Omega=(2\pi)^3/V}$.
Then the Fourier transform becomes
\begin{subequations}
\begin{align}
O(\blR+\blr)  &=\sum_\blG e^{ i\blG\cdot\blr}\,
                 \int_\Omega \frac{d^3\blq}{(2\pi)^3}\, \mathcal{O}(\blq+\blG) e^{i\blq\cdot(\blr+\blR)}
                 \label{Eqn:DFT} \\
\mathcal{O}(\blq+\blG)&= \sum_\blR e^{-i\blq\cdot\blR}\,
                \int_V d^3\blr\, O(\blr+\blR) e^{-i(\blq+\blG)\cdot\blr}
                \label{Eqn:IFT}
\end{align}
\label{Eqn:FTs}
\end{subequations}

\subsubsection{Periodic quantities}
%*********************************

A periodic function $O(\blr)$ is characterized by the property ${O(\blR+\blr)=O(\blr)}$.
Substitution of this identity into Eq.~(\ref{Eqn:IFT}) leads to
\bea
\mathcal{O}(\blq+\blG)&=&\sum_\blR e^{-i\blq\cdot\blR} \int_V d^3\blr\, O(\blr) e^{-i(\blq+\blG)\cdot\blr}   \nonumber \\
            &=&\Omega\d(\blq) \int_V d^3\blr\, O(\blr) e^{-i\blG\cdot\blr} \nonumber \\
            &=&\Omega \d(\blq) \tilde{\mathcal{O}}(\blG),
\eea
i.e., the Fourier transform of a periodic function has non-zero components only for $\blq=\zero$.
We can use this result in Eq.~(\ref{Eqn:DFT}) to obtain 
\bea
O(\blR+\blr)&=& \Omega\sum_\blG e^{ i\blG\cdot\blr} \int_\Omega \frac{d^3\blq}{(2\pi)^3}\, \d(\blq) O(\blG) e^{i\blq\cdot(\blr+\blR)} \nonumber \\
            &=& \frac{1}{V}\sum_\blG e^{ i\blG\cdot\blr} \tilde{\mathcal{O}}(\blG) =O(\blr)
\eea

Thus the Fourier transforms in Eq.~(\ref{Eqn:FTs}) reduce to
\begin{subequations}
\begin{align}
O(\blr) &=\frac{1}{V}\sum_\blG\,\tilde{\mathcal{O}}(\blG)\,e^{ i\blG\cdot\blr} \\
\tilde{\mathcal{O}}(\blG) &=\int_V d^3\blr\,O(\blr)\,e^{-i\blG\cdot\blr}
\end{align}
\end{subequations}
and the macroscopic part of a periodic quantity, i.e., its average over the
unit cell, is $\tilde{\mathcal{O}}(\blG=\zero)/V$.

\subsubsection{Macroscopic quantities}
%***************************************
Here we are interested in processes in which the transferred momentum $\blq+\blG$ is small.
Let us therefore consider a (non periodic) function 
for which, in reciprocal space, only the $\blG=\zero$ component is non-zero, i.e.,
${\mathcal{O}(\blq+\blG)=\d_{\blG,\zero}\overline{\mathcal{O}}(\blq)}$.
Its Fourier transform in real space, given by
\bea
O(\blR+\blr)&=& \sum_\blG e^{ i\blG\cdot\blr} \int_\Omega \frac{d^3\blq}{(2\pi)^3}\, \d_{\blG,\zero} \overline{\mathcal{O}}(\blq) e^{i\blq\cdot(\blr+\blR)}   \nonumber \\
            &=& \int_\Omega \frac{d^3\blq}{(2\pi)^3}\, \overline{\mathcal{O}}(\blq) e^{i\blq\cdot(\blr+\blR)}  
\label{eq:macro_R}
\eea
depends both on $\blR$ and $\blr$, but
with a smooth dependence on $\blr$ since it has no fast oscillating
$\blG$ component.
Its inverse Fourier transform is
\bea
\overline{\mathcal{O}}(\blq) &=& \sum_R e^{-i\blq\cdot\blR} \left( \int_V d^3\blr\, e^{-i\blq\cdot\blr} O(\blR+\blr) \right).
\label{eq:macro_G}
\eea
If $e^{-i\blq\cdot\blr}\simeq 1$ we can neglect the $\blr$ dependence and consider
\be
\overline{O}(\blR)\equiv \frac{1}{V} \int_V d^3\blr\, O(\blR+\blr).
\ee
We thus obtain
\begin{subequations}
\begin{align}
   \overline{O}(\blR) &\simeq\int_\Omega\,\frac{d^3\blq}{(2\pi)^3}\,\overline{\mathcal{O}}(\blq)\,e^{i\blq\cdot\blR} \\
   \overline{\mathcal{O}}(\blq)         &\simeq V\,\sum_R\,\overline{O}(\blR)\,e^{-i\blq\cdot\blR}
\end{align}
\end{subequations}
The relation $e^{-i\blq\cdot\blr}\simeq 1$ holds for small $\blq$, i.e. in the ``macroscopic limit", where
one can consider the volume $V$ as infinitesimal ($V\ra d^3\blR$ and $\W\ra \mathbb{R}^3$)
and $\blR$ becomes a continuous variable. Indeed we can assume $\blq$ or small $V$ if $2\pi/q\gg V^{1/3}$.
In this limit materials are considered as a continuum, their atomistic structure is neglected
and the macroscopic integrals used to describe electromagnetism in classical media are recovered.
Instead for short wavelengths, i.e. $2\pi/q\ll V^{1/3}$, the atomistic
structure can never be neglected and one is forced to use Eqs.~\eqref{eq:macro_R}-\eqref{eq:macro_G}
when computing the response induced at $\blG=\zero$ in PBC.

%********************************************************************************
\section{Full dielectric tensor} \label{App:trans_comp}
%********************************************************************************
In general the dielectric tensor $\tens{\ve}^M(\qq,\w)$
can be decomposed into a longitudinal component, a transverse component, and mixed components as
\be
\tens{\ve}^M(\qq,\w)=
\left( \begin{array}{cc}
\ve^L & \mathbf{\ve}^{LT} \\
\mathbf{\ve}^{TL} & \tens{\ve}^T \\ \end{array} \right).
\ee
For isotropic systems, in particular, there exist only two independent components, i.e., the longitudinal one ($\ve^L$) and transverse one ($\ve^T$), so that the dielectric function reads
\be
\ve^M_{ij}(\qq,\w)=\ve^L(\qq,\omega)\frac{q_iq_j}{|\qq|^2}+\ve^T(\qq,\w)\left(\d_{ij}-\frac{q_iq_j}{|\qq|^2}\right).\\
\ee
The longitudinal component describes the response to longitudinal fields, which are involved, for example, in electron energy-loss experiments, where the scattering cross-section of an electron traversing a medium is proportional to $-Im\left\{1/\ve^{L}(\qq,\w)\right\}$. The transverse component describes the response to optical fields, which are characterized by small $q\approx\omega/c\approx0$. In the long-wavelength limit $q\rightarrow 0$ the two quantities are equal \cite{NozieresPines,Adler1962}, thus optical and energy-loss measurements contain the same physical information. In particular for cubic symmetry we have
\begin{eqnarray}
\lim_{\qq \ra \zero}\tens{\ve}^M(\qq,\w)=\ve^M(\omega)\one.
\label{Eqn:epsilon_jj_cubic}
\end{eqnarray}

Since in the present manuscript we only deal with longitudinal perturbations and longitudinal 
external fields, when we write a scalar dielectric function this is understood to be
its longitudinal component, i.e. ${\ve(\qq,\w)=\ve^L(\qq,\w)}$. The same applies to other quantities
such as the current-current response function ${\x_{jj}(\qq,\w)=\x^L_{\blj\blj}(\qq,\w)=\x_{j^Lj^L}(\qq,\w)}$,
the external potential ${A(\qq,\w)=\blA^L(\qq,\w)}$, the current ${j(\qq,\w)=\blj^L(\qq,\w)}$,
the polarization ${P(\qq,\w)=\blP^L(\qq,\w)}$, and the dipoles ${d_\qq=\bld^L_\qq}$.

%********************************************
\section{Gauges} \label{App:gauges}
%%*******************************************

The scalar and vector potentials $\phi(\blx,t)$ and $\blA(\blx,t)$, respectively, describe a general electromagnetic field by
\begin{eqnarray}
\mathbf{E}(\blx,t) &=-c^{-1}&\partial_t\, \blA(\blx,t)-\nabla \phi(\blx,t), 
\nonumber \\
\mathbf{B}(\blx,t) &= c^{-1}&\nabla\times\blA(\blx,t).
\end{eqnarray}
The electromagnetic field is invariant under the gauge transformations
\begin{eqnarray}
\phi(\blx,t)&\rightarrow&\phi(\blx,t)- c^{-1}\,\partial_t\Lambda(\blx,t),\nonumber\\
\blA(\blx,t)&\rightarrow&\mathbf{A}(\blx,t)+\nabla \Lambda(\blx,t),\nonumber\\
\label{Eqn:gauge_transformation}
\end{eqnarray}
Here $\Lambda(\mathbf{x},t)$ is a differentiable, but, otherwise, arbitrary function of $\mathbf{x}$ and $t$.
Notice that in quantum mechanics the gauge transformation also modifies the wave function phase as:
\begin{equation}
    \psi(\blx,t)\rightarrow\psi(\blx,t)e^{i \frac{\Lambda(\blx,t)}{c}}. \nonumber\\
\end{equation}
One can use the gauge freedom to map a problem in an equivalent one, which is maybe easier to solve.
For example, one can completely gauge transform the scalar potential $\phi$ into a vector potential of the form
\begin{equation}
\blA(\blx,t)=c\int_0^t\mathbf{\nabla} \phi(\blx,t')dt',
\end{equation}
using $\L(\blx,t)=c\int_0^t \phi(\blx,t')dt'$. This is the Weyl gauge.
Such a vector potential, being expressed as the gradient of a scalar, is longitudinal, i.e. it describes a
longitudinal vector field, since its Fourier transform is parallel to $\mathbf{q}$ for any $\mathbf{q}$.
Using $\L(\blr,t)=-\int A^L(\blx',t) d^3\blx'$, one can gauge transform the longitudinal component of
$\blA(\blx,t)$ to a scalar potential as
\begin{equation}
\phi(\blx,t)=c^{-1}\int \partial_t A^L(\blx',t) d^3\blx'.
\end{equation}

This is the Coulomb gauge defined by $\mathbf{\nabla}\cdot\blA(\blx,t)=0$. 
In the dipole approximation, i.e. $\mathbf{E}(\blx,t)\approx \mathbf{E}(\zero,t)$, the two gauges
reduce to the length and the velocity gauge.

Gauge transformation affects only the potentials describing longitudinal fields
as it is clear if Eq.~\eqref{Eqn:gauge_transformation} is written in Fourier space.
Indeed the transverse part of the vector potential is gauge independent and can never be described
in terms of a scalar potential.

%********************************************************************************
\section{Macroscopic response functions\label{App:LR_C}}
%********************************************************************************

In a system with translation invariance symmetry only for $\blx=\blR$,
the longitudinal potentials $\d\phi^{\blG'}(\blq,\omega)$ and $\d A^{\blG'}(\blq,\omega)$ 
in general induce variations at any $\blq+\blG$ component with $\blG$ also different 
from $\blG'$ (we write here the dependence on $\blG$ as a superscript for convenience).
Indeed we can formally write  the component of the induced density and current
density linear in the perturbing potentials of form \eqref{Eqn:PPa}-\eqref{Eqn:PPb}
(using either the length or the velocity gauge) as
\bea
\delta \rho^{\blG}(\blq,\omega)&=&\sum_{\blG'}\chi^{\blG,\blG'}_{\rho\rho}(\blq,\omega)\delta \phi^{\blG'}(\blq,\omega) \nonumber\\
c\, \delta j^{\blG}(\blq,\omega)&=&
\sum_{\blG'} \chi^{\blG,\blG'}_{j^pj^p}(\blq,\omega) \delta A^{\blG'}(\blq,\omega)\nonumber\\
&&+ \frac{1}{V}\sum_{\blG'}\rho^{\blG-\blG'}_0\delta A^{\blG'}(\blq,\omega)
\eea
where $\blq$ is now restricted to the first Brillouin zone, $\blG$ is a reciprocal lattice vector, and 
\bea
\chi_{ab}(\qq,\qq',\w)&=&\frac{1}{(2\pi)^3}\int d\blr e^{-i\qq\cdot\blr}\int d\blr' e^{-i\qq\cdot\blr'} \nonumber\\
&&\times \int_0^\infty d\tau \chi_{ab}(\blr,\blr',\tau).
\eea
If we consider a perturbation with only the $\blG'=\zero$ component,  i.e.
$\d\phi^{\blG'}(\blq,\omega)=\d\phi(\blq,\omega)\,\d_{\blG',\zero}$ and
$\d\blA^{\blG'}(\blq,\omega)=\d\blA(\blq,\omega)\,\d_{\blG',\zero}$
 , and look for the variation of the macroscopic
induced density and current density, i.e. their $\blG=\zero$ Fourier component, 
we arrive at (\ref{Eqn:chi_rr_IP}) and (\ref{Eqn:chi_jj_IP}).

%*********************************************************
\section{intraband contribution to C\label{App:intra}}
%%*********************************************************
Following similar steps as in Ref.~[\onlinecite{Romaniello2005}] the intraband contribution to $\bar{\chi}^{IP}_{dd}$ becomes

\begin{multline}
\x^{IP,\textrm{intra}}_{dd}(\blq,\omega) 
       = \frac{1}{8\pi^3}\sum_{i} \int_{S_i}
   \frac{d^2{\mathbf{k}}}{|\varv^F_{i\kk}|}\,
        j^{p,IP}_{(i\kk+\qq i\kk)}\, j^{p,IP}_{(i\kk i\kk+\qq)}\\
       \times
  \frac{1}{\omega^2}\frac{(\omega/q)^2}{(\varv^F_{i\kk}\cdot\hat{\mathbf{q}})^2-(\omega^+/q)^2},
                 \label{Eqn:chi_dd_IP_intra_2}
\end{multline}
where $\sum_{\kk}$ has been replaced by $V/(2\pi)^3\int d\kk$ and the integration over the $k$-space reduced to an integral over the sheets $S_i$ of the Fermi surface originating by the partially occupied bands $i$. Here $\varv^F_{i\kk}$ is the Fermi velocity. For the frequency-dependent factor we can use the Cauchy theorem and write
 \begin{multline}
\frac{(\omega/q)^2}{(\varv^F_{i\kk}\cdot{\hat{\mathbf{q}}})^2-(\omega^+/q)^2}= 
\mathcal{P}\frac{(\omega/q)^2}{(\varv^F_{i\kk}\cdot{\hat{\mathbf{q}}})^{2}-(\omega/q)^2} \\
+\frac{1}{\omega^2}i\pi({\omega}/{q})^2\big[\d(\varv^F_{i\kk}\cdot{\hat{\mathbf{q}}}-{\omega}/{q})
+\d(\varv^F_{i\kk}\cdot{\hat{\mathbf{q}}}+{\omega}/{q})\big].
\label{Eqn:Ch3_Cauchy1}
\end{multline}

In optical experiments $\omega/q$ is of the order of the velocity of light $c$.
\footnote{The direction of the $\qq\ra\zero$ limit is a delicate point.
Indeed the $(\zero,0)$ point in the $(\qq,\w)$ plane is non analytic, i.e. the value
of the dielectric function depends on the direction of the limit $(\qq,\w)\ra(\zero,0)$.
This direction is determined by the experiment we would like to describe. In optical experiments the direction of interest is $\omega=c\qq$. 
Other directions in the $(\qq,\omega)$ plane may be of interest. 
For example in electron-energy loss experiments one measures the inverse dielectric function 
$\ve^{-1}(\qq,\w)$ at fixed momentum $\qq_{exp}$. The line $\qq=\qq_{exp}$
always crosses the $\varv^F(\qq)$ line, thus a peak must always appear if
we derive $\ve(\qq,\w)$ from the computed electron energy loss function
at $\qq_{exp}$. Therefore for $\qq_{exp}\ra\zero$ we always obtain
the Drude-like tail in the absorption spectrum, also without smearing.}
Therefore, a Drude-like peak in the absorption can be described only if $|\varv^F_{n\kk}|\approx c$.
At the IP level, using real energies, this never happens; therefore the imaginary part in Eq.\ (\ref{Eqn:Ch3_Cauchy1}) is zero and the real part reduces to -1 in the limit of ${q\rightarrow0}$ (and finite $\w$). 
In this case the intraband contribution to the dielectric function is real and reads
\begin{multline}
 \lim_{\blq\rightarrow 0}\ve[\chi^{IP,\textrm{intra}}_{dd}](\blq,\w)=\ve^{IP,\textrm{intra}}(\w) \\
 =\frac{-1}{{\pi^2}\omega^2} \lim_{\blq\rightarrow 0}\sum_{i}\int_{S_i}
   \frac{d^2{\mathbf{k}}}{|\nabla_{\mathbf{k}}\epsilon_{i{\mathbf{k}}}|}\,
  j^{p,IP}_{(i\kk+\qq i\kk)}\, j^{p,IP}_{(i\kk i\kk+\qq)}.
\label{Eqn:e_intra}
\end{multline}

In the excitonic case the treatment of the intraband contribution is more involved and requires
also the Taylor expansion of $E_\lambda(\qq)$ and $A_{\lambda,\qq}$.

%********************************************************************************
\addcontentsline{toc}{chapter}{Bibliography}
\bibliographystyle{apsrev4-1}
\bibliography{manuscript}

\begin{thebibliography}{81}%
\makeatletter
\providecommand \@ifxundefined [1]{%
 \@ifx{#1\undefined}
}%
\providecommand \@ifnum [1]{%
 \ifnum #1\expandafter \@firstoftwo
 \else \expandafter \@secondoftwo
 \fi
}%
\providecommand \@ifx [1]{%
 \ifx #1\expandafter \@firstoftwo
 \else \expandafter \@secondoftwo
 \fi
}%
\providecommand \natexlab [1]{#1}%
\providecommand \enquote  [1]{``#1''}%
\providecommand \bibnamefont  [1]{#1}%
\providecommand \bibfnamefont [1]{#1}%
\providecommand \citenamefont [1]{#1}%
\providecommand \href@noop [0]{\@secondoftwo}%
\providecommand \href [0]{\begingroup \@sanitize@url \@href}%
\providecommand \@href[1]{\@@startlink{#1}\@@href}%
\providecommand \@@href[1]{\endgroup#1\@@endlink}%
\providecommand \@sanitize@url [0]{\catcode `\\12\catcode `\$12\catcode
  `\&12\catcode `\#12\catcode `\^12\catcode `\_12\catcode `\%12\relax}%
\providecommand \@@startlink[1]{}%
\providecommand \@@endlink[0]{}%
\providecommand \url  [0]{\begingroup\@sanitize@url \@url }%
\providecommand \@url [1]{\endgroup\@href {#1}{\urlprefix }}%
\providecommand \urlprefix  [0]{URL }%
\providecommand \Eprint [0]{\href }%
\providecommand \doibase [0]{http://dx.doi.org/}%
\providecommand \selectlanguage [0]{\@gobble}%
\providecommand \bibinfo  [0]{\@secondoftwo}%
\providecommand \bibfield  [0]{\@secondoftwo}%
\providecommand \translation [1]{[#1]}%
\providecommand \BibitemOpen [0]{}%
\providecommand \bibitemStop [0]{}%
\providecommand \bibitemNoStop [0]{.\EOS\space}%
\providecommand \EOS [0]{\spacefactor3000\relax}%
\providecommand \BibitemShut  [1]{\csname bibitem#1\endcsname}%
\let\auto@bib@innerbib\@empty
%</preamble>
\bibitem [{\citenamefont {Strinati}(1988)}]{Strinati1988}%
  \BibitemOpen
  \bibfield  {author} {\bibinfo {author} {\bibfnamefont {G.}~\bibnamefont
  {Strinati}},\ }\href@noop {} {\bibfield  {journal} {\bibinfo  {journal} {Riv.
  Nuovo Cimento}\ }\textbf {\bibinfo {volume} {11}},\ \bibinfo {pages} {1}
  (\bibinfo {year} {1988})}\BibitemShut {NoStop}%
\bibitem [{\citenamefont {Checkelsky}\ \emph {et~al.}(2012)\citenamefont
  {Checkelsky}, \citenamefont {Ye}, \citenamefont {Onose}, \citenamefont
  {Iwasa},\ and\ \citenamefont {Tokura}}]{Checkelsky2012}%
  \BibitemOpen
  \bibfield  {author} {\bibinfo {author} {\bibfnamefont {J.~G.}\ \bibnamefont
  {Checkelsky}}, \bibinfo {author} {\bibfnamefont {J.}~\bibnamefont {Ye}},
  \bibinfo {author} {\bibfnamefont {Y.}~\bibnamefont {Onose}}, \bibinfo
  {author} {\bibfnamefont {Y.}~\bibnamefont {Iwasa}}, \ and\ \bibinfo {author}
  {\bibfnamefont {Y.}~\bibnamefont {Tokura}},\ }\href@noop {} {\bibfield
  {journal} {\bibinfo  {journal} {Nature Physics}\ }\textbf {\bibinfo {volume}
  {8}},\ \bibinfo {pages} {729} (\bibinfo {year} {2012})}\BibitemShut {NoStop}%
\bibitem [{\citenamefont {Chang}\ \emph {et~al.}(2013)\citenamefont {Chang},
  \citenamefont {Zhang}, \citenamefont {Feng}, \citenamefont {Shen},
  \citenamefont {Zhang}, \citenamefont {Guo}, \citenamefont {Li}, \citenamefont
  {Ou}, \citenamefont {Wei}, \citenamefont {Wang} \emph {et~al.}}]{Chang2013}%
  \BibitemOpen
  \bibfield  {author} {\bibinfo {author} {\bibfnamefont {C.-Z.}\ \bibnamefont
  {Chang}}, \bibinfo {author} {\bibfnamefont {J.}~\bibnamefont {Zhang}},
  \bibinfo {author} {\bibfnamefont {X.}~\bibnamefont {Feng}}, \bibinfo {author}
  {\bibfnamefont {J.}~\bibnamefont {Shen}}, \bibinfo {author} {\bibfnamefont
  {Z.}~\bibnamefont {Zhang}}, \bibinfo {author} {\bibfnamefont
  {M.}~\bibnamefont {Guo}}, \bibinfo {author} {\bibfnamefont {K.}~\bibnamefont
  {Li}}, \bibinfo {author} {\bibfnamefont {Y.}~\bibnamefont {Ou}}, \bibinfo
  {author} {\bibfnamefont {P.}~\bibnamefont {Wei}}, \bibinfo {author}
  {\bibfnamefont {L.-L.}\ \bibnamefont {Wang}},  \emph {et~al.},\ }\href@noop
  {} {\bibfield  {journal} {\bibinfo  {journal} {Science}\ }\textbf {\bibinfo
  {volume} {340}},\ \bibinfo {pages} {167} (\bibinfo {year}
  {2013})}\BibitemShut {NoStop}%
\bibitem [{\citenamefont {Kootstra}\ \emph {et~al.}(2000)\citenamefont
  {Kootstra}, \citenamefont {de~Boeij},\ and\ \citenamefont
  {Snijders}}]{Kootstra2000}%
  \BibitemOpen
  \bibfield  {author} {\bibinfo {author} {\bibfnamefont {F.}~\bibnamefont
  {Kootstra}}, \bibinfo {author} {\bibfnamefont {P.~L.}\ \bibnamefont
  {de~Boeij}}, \ and\ \bibinfo {author} {\bibfnamefont {J.~G.}\ \bibnamefont
  {Snijders}},\ }\href {\doibase http://dx.doi.org/10.1063/1.481315} {\bibfield
   {journal} {\bibinfo  {journal} {J. Chem. Phys.}\ }\textbf {\bibinfo {volume}
  {112}},\ \bibinfo {pages} {6517} (\bibinfo {year} {2000})}\BibitemShut
  {NoStop}%
\bibitem [{\citenamefont {Raimbault}\ \emph {et~al.}(2015)\citenamefont
  {Raimbault}, \citenamefont {de~Boeij}, \citenamefont {Romaniello},\ and\
  \citenamefont {Berger}}]{Raimbault2015}%
  \BibitemOpen
  \bibfield  {author} {\bibinfo {author} {\bibfnamefont {N.}~\bibnamefont
  {Raimbault}}, \bibinfo {author} {\bibfnamefont {P.~L.}\ \bibnamefont
  {de~Boeij}}, \bibinfo {author} {\bibfnamefont {P.}~\bibnamefont
  {Romaniello}}, \ and\ \bibinfo {author} {\bibfnamefont {J.~A.}\ \bibnamefont
  {Berger}},\ }\href {\doibase 10.1103/PhysRevLett.114.066404} {\bibfield
  {journal} {\bibinfo  {journal} {Phys. Rev. Lett.}\ }\textbf {\bibinfo
  {volume} {114}},\ \bibinfo {pages} {066404} (\bibinfo {year}
  {2015})}\BibitemShut {NoStop}%
\bibitem [{\citenamefont {Sipe}\ and\ \citenamefont
  {Ghahramani}(1993)}]{Sipe1993}%
  \BibitemOpen
  \bibfield  {author} {\bibinfo {author} {\bibfnamefont {J.~E.}\ \bibnamefont
  {Sipe}}\ and\ \bibinfo {author} {\bibfnamefont {E.}~\bibnamefont
  {Ghahramani}},\ }\href@noop {} {\bibfield  {journal} {\bibinfo  {journal}
  {Phys. Rev. B}\ }\textbf {\bibinfo {volume} {48}},\ \bibinfo {pages} {11705}
  (\bibinfo {year} {1993})}\BibitemShut {NoStop}%
\bibitem [{\citenamefont {Lamb}\ \emph {et~al.}(1987)\citenamefont {Lamb},
  \citenamefont {Schlicher},\ and\ \citenamefont {Scully}}]{Lamb1987}%
  \BibitemOpen
  \bibfield  {author} {\bibinfo {author} {\bibfnamefont {W.~E.}\ \bibnamefont
  {Lamb}}, \bibinfo {author} {\bibfnamefont {R.~R.}\ \bibnamefont {Schlicher}},
  \ and\ \bibinfo {author} {\bibfnamefont {M.~O.}\ \bibnamefont {Scully}},\
  }\href@noop {} {\bibfield  {journal} {\bibinfo  {journal} {Phys. Rev. A}\
  }\textbf {\bibinfo {volume} {36}},\ \bibinfo {pages} {2763} (\bibinfo {year}
  {1987})}\BibitemShut {NoStop}%
\bibitem [{\citenamefont {Sangalli}\ \emph {et~al.}(2012)\citenamefont
  {Sangalli}, \citenamefont {Marini},\ and\ \citenamefont
  {Debernardi}}]{Sangalli2012}%
  \BibitemOpen
  \bibfield  {author} {\bibinfo {author} {\bibfnamefont {D.}~\bibnamefont
  {Sangalli}}, \bibinfo {author} {\bibfnamefont {A.}~\bibnamefont {Marini}}, \
  and\ \bibinfo {author} {\bibfnamefont {A.}~\bibnamefont {Debernardi}},\
  }\href@noop {} {\bibfield  {journal} {\bibinfo  {journal} {Physical Review
  B}\ }\textbf {\bibinfo {volume} {86}},\ \bibinfo {pages} {125139} (\bibinfo
  {year} {2012})}\BibitemShut {NoStop}%
\bibitem [{\citenamefont {Springborg}\ and\ \citenamefont
  {Kirtman}(2008)}]{Springborg2008}%
  \BibitemOpen
  \bibfield  {author} {\bibinfo {author} {\bibfnamefont {M.}~\bibnamefont
  {Springborg}}\ and\ \bibinfo {author} {\bibfnamefont {B.}~\bibnamefont
  {Kirtman}},\ }\href@noop {} {\bibfield  {journal} {\bibinfo  {journal} {Phys.
  Rev. B}\ }\textbf {\bibinfo {volume} {77}},\ \bibinfo {pages} {045102}
  (\bibinfo {year} {2008})}\BibitemShut {NoStop}%
\bibitem [{\citenamefont {Schafer}\ and\ \citenamefont
  {Wegener}(2002)}]{Schafer}%
  \BibitemOpen
  \bibfield  {author} {\bibinfo {author} {\bibfnamefont {W.}~\bibnamefont
  {Schafer}}\ and\ \bibinfo {author} {\bibfnamefont {M.}~\bibnamefont
  {Wegener}},\ }\href@noop {} {\emph {\bibinfo {title} {Semiconductor Optics
  and Transport Phenomena: From Fundamentals to Current Topics}}}\ (\bibinfo
  {publisher} {Springer},\ \bibinfo {year} {2002})\BibitemShut {NoStop}%
\bibitem [{Note1()}]{Note1}%
  \BibitemOpen
  \bibinfo {note} {The differences out of resonance can be of small importance
  in the linear regime but they become crucial for any non-linear phenomena
  because they play an important role in the construction of high order
  response functions.}\BibitemShut {Stop}%
\bibitem [{\citenamefont {Fetter}\ and\ \citenamefont
  {Walecka}(2003)}]{Fetter2003}%
  \BibitemOpen
  \bibfield  {author} {\bibinfo {author} {\bibfnamefont {A.~L.}\ \bibnamefont
  {Fetter}}\ and\ \bibinfo {author} {\bibfnamefont {J.~D.}\ \bibnamefont
  {Walecka}},\ }\href@noop {} {\emph {\bibinfo {title} {Quantum theory of
  many-particle systems}}}\ (\bibinfo  {publisher} {Courier Corporation},\
  \bibinfo {year} {2003})\BibitemShut {NoStop}%
\bibitem [{\citenamefont {Rohlfing}\ and\ \citenamefont
  {Louie}(2000)}]{Louie2000}%
  \BibitemOpen
  \bibfield  {author} {\bibinfo {author} {\bibfnamefont {M.}~\bibnamefont
  {Rohlfing}}\ and\ \bibinfo {author} {\bibfnamefont {S.~G.}\ \bibnamefont
  {Louie}},\ }\href@noop {} {\bibfield  {journal} {\bibinfo  {journal}
  {Physical Review B}\ }\textbf {\bibinfo {volume} {62}},\ \bibinfo {pages}
  {4927} (\bibinfo {year} {2000})}\BibitemShut {NoStop}%
\bibitem [{\citenamefont {Del~Sole}\ and\ \citenamefont
  {Fiorino}(1984)}]{DelSole1984}%
  \BibitemOpen
  \bibfield  {author} {\bibinfo {author} {\bibfnamefont {R.}~\bibnamefont
  {Del~Sole}}\ and\ \bibinfo {author} {\bibfnamefont {E.}~\bibnamefont
  {Fiorino}},\ }\href@noop {} {\bibfield  {journal} {\bibinfo  {journal}
  {Physical Review B}\ }\textbf {\bibinfo {volume} {29}},\ \bibinfo {pages}
  {4631} (\bibinfo {year} {1984})}\BibitemShut {NoStop}%
\bibitem [{\citenamefont {Romaniello}\ and\ \citenamefont
  {de~Boeij}(2005)}]{Romaniello2005}%
  \BibitemOpen
  \bibfield  {author} {\bibinfo {author} {\bibfnamefont {P.}~\bibnamefont
  {Romaniello}}\ and\ \bibinfo {author} {\bibfnamefont {P.~L.}\ \bibnamefont
  {de~Boeij}},\ }\href {\doibase 10.1103/PhysRevB.71.155108} {\bibfield
  {journal} {\bibinfo  {journal} {Phys. Rev. B}\ }\textbf {\bibinfo {volume}
  {71}},\ \bibinfo {pages} {155108} (\bibinfo {year} {2005})}\BibitemShut
  {NoStop}%
\bibitem [{\citenamefont {Berger}\ \emph {et~al.}(2005)\citenamefont {Berger},
  \citenamefont {de~Boeij},\ and\ \citenamefont {van Leeuwen}}]{Berger2005}%
  \BibitemOpen
  \bibfield  {author} {\bibinfo {author} {\bibfnamefont {J.~A.}\ \bibnamefont
  {Berger}}, \bibinfo {author} {\bibfnamefont {P.~L.}\ \bibnamefont
  {de~Boeij}}, \ and\ \bibinfo {author} {\bibfnamefont {R.}~\bibnamefont {van
  Leeuwen}},\ }\href {\doibase 10.1103/PhysRevB.71.155104} {\bibfield
  {journal} {\bibinfo  {journal} {Phys. Rev. B}\ }\textbf {\bibinfo {volume}
  {71}},\ \bibinfo {pages} {155104} (\bibinfo {year} {2005})}\BibitemShut
  {NoStop}%
\bibitem [{\citenamefont {Berger}\ \emph {et~al.}(2006)\citenamefont {Berger},
  \citenamefont {Romaniello}, \citenamefont {van Leeuwen},\ and\ \citenamefont
  {de~Boeij}}]{Berger2006}%
  \BibitemOpen
  \bibfield  {author} {\bibinfo {author} {\bibfnamefont {J.~A.}\ \bibnamefont
  {Berger}}, \bibinfo {author} {\bibfnamefont {P.}~\bibnamefont {Romaniello}},
  \bibinfo {author} {\bibfnamefont {R.}~\bibnamefont {van Leeuwen}}, \ and\
  \bibinfo {author} {\bibfnamefont {P.~L.}\ \bibnamefont {de~Boeij}},\ }\href
  {\doibase 10.1103/PhysRevB.74.245117} {\bibfield  {journal} {\bibinfo
  {journal} {Phys. Rev. B}\ }\textbf {\bibinfo {volume} {74}},\ \bibinfo
  {pages} {245117} (\bibinfo {year} {2006})}\BibitemShut {NoStop}%
\bibitem [{\citenamefont {Attaccalite}\ \emph {et~al.}(2011)\citenamefont
  {Attaccalite}, \citenamefont {Gr\"uning},\ and\ \citenamefont
  {Marini}}]{Attaccalite2011}%
  \BibitemOpen
  \bibfield  {author} {\bibinfo {author} {\bibfnamefont {C.}~\bibnamefont
  {Attaccalite}}, \bibinfo {author} {\bibfnamefont {M.}~\bibnamefont
  {Gr\"uning}}, \ and\ \bibinfo {author} {\bibfnamefont {A.}~\bibnamefont
  {Marini}},\ }\href {\doibase 10.1103/PhysRevB.84.245110} {\bibfield
  {journal} {\bibinfo  {journal} {Phys. Rev. B}\ }\textbf {\bibinfo {volume}
  {84}},\ \bibinfo {pages} {245110} (\bibinfo {year} {2011})}\BibitemShut
  {NoStop}%
\bibitem [{\citenamefont {Castro}\ and\ \citenamefont
  {Appel}(2006)}]{Castro2006}%
  \BibitemOpen
  \bibfield  {author} {\bibinfo {author} {\bibfnamefont {A.}~\bibnamefont
  {Castro}}\ and\ \bibinfo {author} {\bibfnamefont {H.~a.}\ \bibnamefont
  {Appel}},\ }\href {\doibase 10.1002/pssb.200642067} {\bibfield  {journal}
  {\bibinfo  {journal} {Phys. Stat. Sol. B}\ }\textbf {\bibinfo {volume}
  {243}},\ \bibinfo {pages} {2465} (\bibinfo {year} {2006})}\BibitemShut
  {NoStop}%
\bibitem [{\citenamefont {Bertsch}\ \emph {et~al.}(2000)\citenamefont
  {Bertsch}, \citenamefont {Iwata}, \citenamefont {Rubio},\ and\ \citenamefont
  {Yabana}}]{Bertsch2000}%
  \BibitemOpen
  \bibfield  {author} {\bibinfo {author} {\bibfnamefont {G.~F.}\ \bibnamefont
  {Bertsch}}, \bibinfo {author} {\bibfnamefont {J.-I.}\ \bibnamefont {Iwata}},
  \bibinfo {author} {\bibfnamefont {A.}~\bibnamefont {Rubio}}, \ and\ \bibinfo
  {author} {\bibfnamefont {K.}~\bibnamefont {Yabana}},\ }\href@noop {}
  {\bibfield  {journal} {\bibinfo  {journal} {Physical Review B}\ }\textbf
  {\bibinfo {volume} {62}},\ \bibinfo {pages} {7998} (\bibinfo {year}
  {2000})}\BibitemShut {NoStop}%
\bibitem [{Note2()}]{Note2}%
  \BibitemOpen
  \bibinfo {note} {In condensed matter the IP approximations is often called
  RPA to be distinguished from RPA plus local field effects.}\BibitemShut
  {Stop}%
\bibitem [{\citenamefont {Wiser}(1963)}]{Wiser1963}%
  \BibitemOpen
  \bibfield  {author} {\bibinfo {author} {\bibfnamefont {N.}~\bibnamefont
  {Wiser}},\ }\href {\doibase 10.1103/PhysRev.129.62} {\bibfield  {journal}
  {\bibinfo  {journal} {Phys. Rev.}\ }\textbf {\bibinfo {volume} {129}},\
  \bibinfo {pages} {62} (\bibinfo {year} {1963})}\BibitemShut {NoStop}%
\bibitem [{\citenamefont {Landau}\ and\ \citenamefont
  {Lifshits}(1965)}]{landau1965quantum}%
  \BibitemOpen
  \bibfield  {author} {\bibinfo {author} {\bibfnamefont {L.~D.}\ \bibnamefont
  {Landau}}\ and\ \bibinfo {author} {\bibfnamefont {E.~M.}\ \bibnamefont
  {Lifshits}},\ }\href@noop {} {\emph {\bibinfo {title} {Quantum Mechanics
  Non-relativistic Theory: Transl. from the Russian by JB Sykes and JS bell. 2d
  Ed., rev. and Enl}}}\ (\bibinfo  {publisher} {Pergamon press},\ \bibinfo
  {year} {1965})\BibitemShut {NoStop}%
\bibitem [{\citenamefont {Giuliani}\ and\ \citenamefont
  {Vignale}(2005)}]{Giuliani}%
  \BibitemOpen
  \bibfield  {author} {\bibinfo {author} {\bibfnamefont {G.}~\bibnamefont
  {Giuliani}}\ and\ \bibinfo {author} {\bibfnamefont {G.}~\bibnamefont
  {Vignale}},\ }\href@noop {} {\emph {\bibinfo {title} {Quantum theory of the
  electron liquid}}}\ (\bibinfo  {publisher} {Cambridge university press},\
  \bibinfo {year} {2005})\BibitemShut {NoStop}%
\bibitem [{Note3()}]{Note3}%
  \BibitemOpen
  \bibinfo {note} {The inclusion of $\protect \sqrt {\Delta f_{mn}({\protect
  \bf q})}$ in the definition of the matrix elements of Eq.~\protect \textup
  {\hbox {\mathsurround \z@ \protect \normalfont (\ignorespaces \ref
  {Eqn:IP_operators}\unskip \@@italiccorr )}} allows to define in the following
  sections (where the the case of interacting electron-hole pairs is
  considered) an excitonic matrix which remains Hermitian also in the general
  case of fractional occupation numbers. See also Ref.~\protect \rev@citealpnum
  {Sangalli2016}.}\BibitemShut {Stop}%
\bibitem [{Note4()}]{Note4}%
  \BibitemOpen
  \bibinfo {note} {Which can be obtained by Taylor expanding ${F(\omega )}$
  around ${\omega =0}$ up to the first order. The last term on the right-hand
  side is then defined from the difference ${F(\omega )-F(0)-F'(0)\protect
  \tmspace +\thinmuskip {.1667em}\omega }$.}\BibitemShut {Stop}%
\bibitem [{Note5()}]{Note5}%
  \BibitemOpen
  \bibinfo {note} {Eq.~\protect \textup {\hbox {\mathsurround \z@ \protect
  \normalfont (\ignorespaces \ref {Eqn:eps_jj_rr_relation}\unskip \@@italiccorr
  )}} implies $q^2 K^{j^pj^p}_{nm,{\protect \bf q}}=\Delta \epsilon
  _{nm}^2({\protect \bf q}) K^{\rho \rho }_{nm,{\protect \bf q}}$, which is
  similar, but not the same, to Eq.~\protect \textup {\hbox {\mathsurround \z@
  \protect \normalfont (\ignorespaces \ref {Eqn:Xjj_Xrr_relation}\unskip
  \@@italiccorr )}}. Notice the non trivial replacement of $\omega
  ^2\rightarrow \Delta \epsilon _{nm}^2({\protect \bf q})$.}\BibitemShut
  {Stop}%
\bibitem [{Note6()}]{Note6}%
  \BibitemOpen
  \bibinfo {note} {The position operator is ill-defined when periodic boundary
  conditions (PBC) are imposed. Here we implicitly work in the crystal momentum
  representation in which the matrix elements of the position operator are
  redefined consistently with the PBC~\cite {Blount1962,Souza2004}}\BibitemShut
  {NoStop}%
\bibitem [{Note7()}]{Note7}%
  \BibitemOpen
  \bibinfo {note} {Here $d_{{\protect \mathbf q}}$ at ${\protect \bf
  q}={\protect \bf 0}$ is the longitudinal dipole, defined by the direction of
  the ${\protect \bf q}\rightarrow {\protect \bf 0}$. See also App.~\ref
  {App:trans_comp}.}\BibitemShut {Stop}%
\bibitem [{Note8()}]{Note8}%
  \BibitemOpen
  \bibinfo {note} {For finite ${\protect \bf q}$ Eq.~\protect \textup {\hbox
  {\mathsurround \z@ \protect \normalfont (\ignorespaces \ref
  {Eqn:eps_dd}\unskip \@@italiccorr )}} is exact to first order in ${\protect
  \bf q}$}\BibitemShut {NoStop}%
\bibitem [{Note9()}]{Note9}%
  \BibitemOpen
  \bibinfo {note} {Here $B^\protect \textrm {IP,inter}=0$ because we are
  considering the longitudinal term only. Indeed the mixed
  longitudinal-transverse terms can instead be different from zero and describe
  the Anomalous Hall effect~\cite {Sangalli2012}.}\BibitemShut {Stop}%
\bibitem [{\citenamefont {Marini}(2001)}]{MariniPhD}%
  \BibitemOpen
  \bibfield  {author} {\bibinfo {author} {\bibfnamefont {A.}~\bibnamefont
  {Marini}},\ }\emph {\bibinfo {title} {Optical and electronic properties of
  Copper and Silver: from Density Functional Theory to Many Body Effects}},\
  \href@noop {} {Ph.D. thesis},\ \bibinfo  {school} {Universit\'a di tor
  vergata, Roma (Italy)} (\bibinfo {year} {2001})\BibitemShut {NoStop}%
\bibitem [{\citenamefont {Bussi}(2004)}]{Bussi2004}%
  \BibitemOpen
  \bibfield  {author} {\bibinfo {author} {\bibfnamefont {G.}~\bibnamefont
  {Bussi}},\ }\href@noop {} {\bibfield  {journal} {\bibinfo  {journal} {Physica
  Scripta}\ }\textbf {\bibinfo {volume} {2004}},\ \bibinfo {pages} {141}
  (\bibinfo {year} {2004})}\BibitemShut {NoStop}%
\bibitem [{Note10()}]{Note10}%
  \BibitemOpen
  \bibinfo {note} {As commonly done in the literature, we neglect the term
  $i\delta W/\delta G$ in the kernel.}\BibitemShut {Stop}%
\bibitem [{Note11()}]{Note11}%
  \BibitemOpen
  \bibinfo {note} {For a detailed treatment of the space indexes at finite
  momentum see Ref.~[\protect \rev@citealpnum {Sottile2013}]; for the treatment
  of occupations factors see Ref.~[\protect \rev@citealpnum
  {Sangalli2016}])}\BibitemShut {NoStop}%
\bibitem [{\citenamefont {Onida}\ \emph {et~al.}(2002)\citenamefont {Onida},
  \citenamefont {Reining},\ and\ \citenamefont {Rubio}}]{Onida2002}%
  \BibitemOpen
  \bibfield  {author} {\bibinfo {author} {\bibfnamefont {G.}~\bibnamefont
  {Onida}}, \bibinfo {author} {\bibfnamefont {L.}~\bibnamefont {Reining}}, \
  and\ \bibinfo {author} {\bibfnamefont {A.}~\bibnamefont {Rubio}},\
  }\href@noop {} {\bibfield  {journal} {\bibinfo  {journal} {Rev. Mod. Phys.}\
  }\textbf {\bibinfo {volume} {74}},\ \bibinfo {pages} {601} (\bibinfo {year}
  {2002})}\BibitemShut {NoStop}%
\bibitem [{\citenamefont {Gatti}\ and\ \citenamefont
  {Sottile}(2013)}]{Sottile2013}%
  \BibitemOpen
  \bibfield  {author} {\bibinfo {author} {\bibfnamefont {M.}~\bibnamefont
  {Gatti}}\ and\ \bibinfo {author} {\bibfnamefont {F.}~\bibnamefont
  {Sottile}},\ }\href {\doibase 10.1103/PhysRevB.88.155113} {\bibfield
  {journal} {\bibinfo  {journal} {Phys. Rev. B}\ }\textbf {\bibinfo {volume}
  {88}},\ \bibinfo {pages} {155113} (\bibinfo {year} {2013})}\BibitemShut
  {NoStop}%
\bibitem [{Note12()}]{Note12}%
  \BibitemOpen
  \bibinfo {note} {The exact Dyson equations holds only for the time-ordered
  two-particles propagator, which is formally derived assuming zero temperature
  (i.e. integer occupation numbers also in case of metals). To consider
  fractional occupations one would need to introduce a finite temperature
  formalism. However, using a static kernel, the Dyson equation at finite
  temperature reduces to a Dyson equation identical to Eq.~\protect \textup
  {\hbox {\mathsurround \z@ \protect \normalfont (\ignorespaces \ref
  {Eqn:L(w_static)}\unskip \@@italiccorr )}} but for the retarded propagator,
  which is indeed what is needed to define the dielectric function. Thus from
  now on we can consider all quantities as retarded functions and forget about
  the time-ordered formalism.}\BibitemShut {Stop}%
\bibitem [{Note13()}]{Note13}%
  \BibitemOpen
  \bibinfo {note} {We assume that zero energy transitions at the BSE level
  originates from zero energy transitions at the IP level. Zero energy poles at
  the BSE level may originate from finite energies transition at the IP level
  as well. However this case would point to an instability of the ground state
  which would be degenerate to an excited state. We exclude this possibility in
  the present work.}\BibitemShut {Stop}%
\bibitem [{\citenamefont {Giannozzi}\ \emph {et~al.}(2009)\citenamefont
  {Giannozzi}, \citenamefont {Baroni}, \citenamefont {Bonini}, \citenamefont
  {Calandra}, \citenamefont {Car}, \citenamefont {Cavazzoni}, \citenamefont
  {Ceresoli}, \citenamefont {Chiarotti}, \citenamefont {Cococcioni},
  \citenamefont {Dabo}, \citenamefont {Corso}, \citenamefont {de~Gironcoli},
  \citenamefont {Fabris}, \citenamefont {Fratesi}, \citenamefont {Gebauer},
  \citenamefont {Gerstmann}, \citenamefont {Gougoussis}, \citenamefont
  {Kokalj}, \citenamefont {Lazzeri}, \citenamefont {Martin-Samos},
  \citenamefont {Marzari}, \citenamefont {Mauri}, \citenamefont {Mazzarello},
  \citenamefont {Paolini}, \citenamefont {Pasquarello}, \citenamefont
  {Paulatto}, \citenamefont {Sbraccia}, \citenamefont {Scandolo}, \citenamefont
  {Sclauzero}, \citenamefont {Seitsonen}, \citenamefont {Smogunov},
  \citenamefont {Umari},\ and\ \citenamefont {Wentzcovitch}}]{Giannozzi2009}%
  \BibitemOpen
  \bibfield  {author} {\bibinfo {author} {\bibfnamefont {P.}~\bibnamefont
  {Giannozzi}}, \bibinfo {author} {\bibfnamefont {S.}~\bibnamefont {Baroni}},
  \bibinfo {author} {\bibfnamefont {N.}~\bibnamefont {Bonini}}, \bibinfo
  {author} {\bibfnamefont {M.}~\bibnamefont {Calandra}}, \bibinfo {author}
  {\bibfnamefont {R.}~\bibnamefont {Car}}, \bibinfo {author} {\bibfnamefont
  {C.}~\bibnamefont {Cavazzoni}}, \bibinfo {author} {\bibfnamefont
  {D.}~\bibnamefont {Ceresoli}}, \bibinfo {author} {\bibfnamefont {G.~L.}\
  \bibnamefont {Chiarotti}}, \bibinfo {author} {\bibfnamefont {M.}~\bibnamefont
  {Cococcioni}}, \bibinfo {author} {\bibfnamefont {I.}~\bibnamefont {Dabo}},
  \bibinfo {author} {\bibfnamefont {A.~D.}\ \bibnamefont {Corso}}, \bibinfo
  {author} {\bibfnamefont {S.}~\bibnamefont {de~Gironcoli}}, \bibinfo {author}
  {\bibfnamefont {S.}~\bibnamefont {Fabris}}, \bibinfo {author} {\bibfnamefont
  {G.}~\bibnamefont {Fratesi}}, \bibinfo {author} {\bibfnamefont
  {R.}~\bibnamefont {Gebauer}}, \bibinfo {author} {\bibfnamefont
  {U.}~\bibnamefont {Gerstmann}}, \bibinfo {author} {\bibfnamefont
  {C.}~\bibnamefont {Gougoussis}}, \bibinfo {author} {\bibfnamefont
  {A.}~\bibnamefont {Kokalj}}, \bibinfo {author} {\bibfnamefont
  {M.}~\bibnamefont {Lazzeri}}, \bibinfo {author} {\bibfnamefont
  {L.}~\bibnamefont {Martin-Samos}}, \bibinfo {author} {\bibfnamefont
  {N.}~\bibnamefont {Marzari}}, \bibinfo {author} {\bibfnamefont
  {F.}~\bibnamefont {Mauri}}, \bibinfo {author} {\bibfnamefont
  {R.}~\bibnamefont {Mazzarello}}, \bibinfo {author} {\bibfnamefont
  {S.}~\bibnamefont {Paolini}}, \bibinfo {author} {\bibfnamefont
  {A.}~\bibnamefont {Pasquarello}}, \bibinfo {author} {\bibfnamefont
  {L.}~\bibnamefont {Paulatto}}, \bibinfo {author} {\bibfnamefont
  {C.}~\bibnamefont {Sbraccia}}, \bibinfo {author} {\bibfnamefont
  {S.}~\bibnamefont {Scandolo}}, \bibinfo {author} {\bibfnamefont
  {G.}~\bibnamefont {Sclauzero}}, \bibinfo {author} {\bibfnamefont {A.~P.}\
  \bibnamefont {Seitsonen}}, \bibinfo {author} {\bibfnamefont {A.}~\bibnamefont
  {Smogunov}}, \bibinfo {author} {\bibfnamefont {P.}~\bibnamefont {Umari}}, \
  and\ \bibinfo {author} {\bibfnamefont {R.~M.}\ \bibnamefont {Wentzcovitch}},\
  }\href {http://stacks.iop.org/0953-8984/21/i=39/a=395502} {\bibfield
  {journal} {\bibinfo  {journal} {J. Phys.: Condensed Matter}\ }\textbf
  {\bibinfo {volume} {21}},\ \bibinfo {pages} {395502} (\bibinfo {year}
  {2009})}\BibitemShut {NoStop}%
\bibitem [{\citenamefont {Gonze}\ \emph {et~al.}(2002)\citenamefont {Gonze}
  \emph {et~al.}}]{abinit}%
  \BibitemOpen
  \bibfield  {author} {\bibinfo {author} {\bibfnamefont {X.}~\bibnamefont
  {Gonze}} \emph {et~al.},\ }\href@noop {} {\bibfield  {journal} {\bibinfo
  {journal} {Comput. Mater. Sci.}\ }\textbf {\bibinfo {volume} {25}},\ \bibinfo
  {pages} {478 } (\bibinfo {year} {2002})}\BibitemShut {NoStop}%
\bibitem [{\citenamefont {Goedecker}\ \emph {et~al.}(1996)\citenamefont
  {Goedecker}, \citenamefont {Teter},\ and\ \citenamefont
  {Hutter}}]{Goedecker1996}%
  \BibitemOpen
  \bibfield  {author} {\bibinfo {author} {\bibfnamefont {S.}~\bibnamefont
  {Goedecker}}, \bibinfo {author} {\bibfnamefont {M.}~\bibnamefont {Teter}}, \
  and\ \bibinfo {author} {\bibfnamefont {J.}~\bibnamefont {Hutter}},\ }\href
  {\doibase 10.1103/PhysRevB.54.1703} {\bibfield  {journal} {\bibinfo
  {journal} {Phys. Rev. B}\ }\textbf {\bibinfo {volume} {54}},\ \bibinfo
  {pages} {1703} (\bibinfo {year} {1996})}\BibitemShut {NoStop}%
\bibitem [{\citenamefont {Marini}\ \emph {et~al.}(2009)\citenamefont {Marini},
  \citenamefont {Hogan}, \citenamefont {Gr\"uning},\ and\ \citenamefont
  {Varsano}}]{Marini2009}%
  \BibitemOpen
  \bibfield  {author} {\bibinfo {author} {\bibfnamefont {A.}~\bibnamefont
  {Marini}}, \bibinfo {author} {\bibfnamefont {C.}~\bibnamefont {Hogan}},
  \bibinfo {author} {\bibfnamefont {M.}~\bibnamefont {Gr\"uning}}, \ and\
  \bibinfo {author} {\bibfnamefont {D.}~\bibnamefont {Varsano}},\ }\href@noop
  {} {\bibfield  {journal} {\bibinfo  {journal} {Computer Physics
  Communications}\ }\textbf {\bibinfo {volume} {180}},\ \bibinfo {pages} {1392
  } (\bibinfo {year} {2009})}\BibitemShut {NoStop}%
\bibitem [{Note14()}]{Note14}%
  \BibitemOpen
  \bibinfo {note} {Yambo standard implementation uses a density-based
  approach}\BibitemShut {NoStop}%
\bibitem [{\citenamefont {Giustino}(2016)}]{Giustino2016}%
  \BibitemOpen
  \bibfield  {author} {\bibinfo {author} {\bibfnamefont {F.}~\bibnamefont
  {Giustino}},\ }\href@noop {} {\bibfield  {journal} {\bibinfo  {journal}
  {arXiv preprint arXiv:1603.06965}\ } (\bibinfo {year} {2016})}\BibitemShut
  {NoStop}%
\bibitem [{\citenamefont {Bernardi}\ \emph {et~al.}(2014)\citenamefont
  {Bernardi}, \citenamefont {Vigil-Fowler}, \citenamefont {Lischner},
  \citenamefont {Neaton},\ and\ \citenamefont {Louie}}]{Bernardi2014}%
  \BibitemOpen
  \bibfield  {author} {\bibinfo {author} {\bibfnamefont {M.}~\bibnamefont
  {Bernardi}}, \bibinfo {author} {\bibfnamefont {D.}~\bibnamefont
  {Vigil-Fowler}}, \bibinfo {author} {\bibfnamefont {J.}~\bibnamefont
  {Lischner}}, \bibinfo {author} {\bibfnamefont {J.~B.}\ \bibnamefont
  {Neaton}}, \ and\ \bibinfo {author} {\bibfnamefont {S.~G.}\ \bibnamefont
  {Louie}},\ }\href {\doibase 10.1103/PhysRevLett.112.257402} {\bibfield
  {journal} {\bibinfo  {journal} {Phys. Rev. Lett.}\ }\textbf {\bibinfo
  {volume} {112}},\ \bibinfo {pages} {257402} (\bibinfo {year}
  {2014})}\BibitemShut {NoStop}%
\bibitem [{\citenamefont {Aryasetiawan}\ and\ \citenamefont
  {Gunnarsson}(1998)}]{Aryasetiawan1998}%
  \BibitemOpen
  \bibfield  {author} {\bibinfo {author} {\bibfnamefont {F.}~\bibnamefont
  {Aryasetiawan}}\ and\ \bibinfo {author} {\bibfnamefont {O.}~\bibnamefont
  {Gunnarsson}},\ }\href@noop {} {\bibfield  {journal} {\bibinfo  {journal}
  {Reports on Progress in Physics}\ }\textbf {\bibinfo {volume} {61}},\
  \bibinfo {pages} {237} (\bibinfo {year} {1998})}\BibitemShut {NoStop}%
\bibitem [{\citenamefont {Cazzaniga}\ \emph {et~al.}(2010)\citenamefont
  {Cazzaniga}, \citenamefont {Caramella}, \citenamefont {Manini},\ and\
  \citenamefont {Onida}}]{Cazzaniga2010}%
  \BibitemOpen
  \bibfield  {author} {\bibinfo {author} {\bibfnamefont {M.}~\bibnamefont
  {Cazzaniga}}, \bibinfo {author} {\bibfnamefont {L.}~\bibnamefont
  {Caramella}}, \bibinfo {author} {\bibfnamefont {N.}~\bibnamefont {Manini}}, \
  and\ \bibinfo {author} {\bibfnamefont {G.}~\bibnamefont {Onida}},\ }\href
  {\doibase 10.1103/PhysRevB.82.035104} {\bibfield  {journal} {\bibinfo
  {journal} {Phys. Rev. B}\ }\textbf {\bibinfo {volume} {82}},\ \bibinfo
  {pages} {035104} (\bibinfo {year} {2010})}\BibitemShut {NoStop}%
\bibitem [{\citenamefont {Dold}\ and\ \citenamefont {Mecke}(1975)}]{Dold1975}%
  \BibitemOpen
  \bibfield  {author} {\bibinfo {author} {\bibfnamefont {B.}~\bibnamefont
  {Dold}}\ and\ \bibinfo {author} {\bibfnamefont {R.}~\bibnamefont {Mecke}},\
  }\href@noop {} {\bibfield  {journal} {\bibinfo  {journal} {Optik Am.}\
  }\textbf {\bibinfo {volume} {65}},\ \bibinfo {pages} {742} (\bibinfo {year}
  {1975})}\BibitemShut {NoStop}%
\bibitem [{\citenamefont {Ehrenreich}\ and\ \citenamefont
  {Philipp}(1962)}]{ehrenreich}%
  \BibitemOpen
  \bibfield  {author} {\bibinfo {author} {\bibfnamefont {H.}~\bibnamefont
  {Ehrenreich}}\ and\ \bibinfo {author} {\bibfnamefont {H.~R.}\ \bibnamefont
  {Philipp}},\ }\href {\doibase 10.1103/PhysRev.128.1622} {\bibfield  {journal}
  {\bibinfo  {journal} {Phys. Rev.}\ }\textbf {\bibinfo {volume} {128}},\
  \bibinfo {pages} {1622} (\bibinfo {year} {1962})}\BibitemShut {NoStop}%
\bibitem [{\citenamefont {Stahrenberg}\ \emph {et~al.}(2001)\citenamefont
  {Stahrenberg}, \citenamefont {Herrmann}, \citenamefont {Wilmers},
  \citenamefont {Esser}, \citenamefont {Richter},\ and\ \citenamefont
  {Lee}}]{stahrenberg}%
  \BibitemOpen
  \bibfield  {author} {\bibinfo {author} {\bibfnamefont {K.}~\bibnamefont
  {Stahrenberg}}, \bibinfo {author} {\bibfnamefont {T.}~\bibnamefont
  {Herrmann}}, \bibinfo {author} {\bibfnamefont {K.}~\bibnamefont {Wilmers}},
  \bibinfo {author} {\bibfnamefont {N.}~\bibnamefont {Esser}}, \bibinfo
  {author} {\bibfnamefont {W.}~\bibnamefont {Richter}}, \ and\ \bibinfo
  {author} {\bibfnamefont {M.~J.~G.}\ \bibnamefont {Lee}},\ }\href {\doibase
  10.1103/PhysRevB.64.115111} {\bibfield  {journal} {\bibinfo  {journal} {Phys.
  Rev. B}\ }\textbf {\bibinfo {volume} {64}},\ \bibinfo {pages} {115111}
  (\bibinfo {year} {2001})}\BibitemShut {NoStop}%
\bibitem [{\citenamefont {Bl{\"o}chl}\ \emph {et~al.}(1994)\citenamefont
  {Bl{\"o}chl}, \citenamefont {Jepsen},\ and\ \citenamefont
  {Andersen}}]{Blochl1994}%
  \BibitemOpen
  \bibfield  {author} {\bibinfo {author} {\bibfnamefont {P.~E.}\ \bibnamefont
  {Bl{\"o}chl}}, \bibinfo {author} {\bibfnamefont {O.}~\bibnamefont {Jepsen}},
  \ and\ \bibinfo {author} {\bibfnamefont {O.~K.}\ \bibnamefont {Andersen}},\
  }\href@noop {} {\bibfield  {journal} {\bibinfo  {journal} {Physical Review
  B}\ }\textbf {\bibinfo {volume} {49}},\ \bibinfo {pages} {16223} (\bibinfo
  {year} {1994})}\BibitemShut {NoStop}%
\bibitem [{\citenamefont {Marini}(2008)}]{Marini2008}%
  \BibitemOpen
  \bibfield  {author} {\bibinfo {author} {\bibfnamefont {A.}~\bibnamefont
  {Marini}},\ }\href {\doibase 10.1103/PhysRevLett.101.106405} {\bibfield
  {journal} {\bibinfo  {journal} {Phys. Rev. Lett.}\ }\textbf {\bibinfo
  {volume} {101}},\ \bibinfo {pages} {106405} (\bibinfo {year}
  {2008})}\BibitemShut {NoStop}%
\bibitem [{\citenamefont {Del~Sole}\ and\ \citenamefont
  {Girlanda}(1993)}]{DelSole1993}%
  \BibitemOpen
  \bibfield  {author} {\bibinfo {author} {\bibfnamefont {R.}~\bibnamefont
  {Del~Sole}}\ and\ \bibinfo {author} {\bibfnamefont {R.}~\bibnamefont
  {Girlanda}},\ }\href {\doibase 10.1103/PhysRevB.48.11789} {\bibfield
  {journal} {\bibinfo  {journal} {Phys. Rev. B}\ }\textbf {\bibinfo {volume}
  {48}},\ \bibinfo {pages} {11789} (\bibinfo {year} {1993})}\BibitemShut
  {NoStop}%
\bibitem [{Note15()}]{Note15}%
  \BibitemOpen
  \bibinfo {note} {Ref.~\protect \rev@citealpnum {Tokman2009} discusses the
  connection between the time derivative of the dipole operator and the
  velocity operator in case of a dephasing of the polarization (i.e. an
  imaginary term in the Hamiltonian) is considered.}\BibitemShut {Stop}%
\bibitem [{\citenamefont {Spataru}\ \emph {et~al.}(2005)\citenamefont
  {Spataru}, \citenamefont {Ismail-Beigi}, \citenamefont {Capaz},\ and\
  \citenamefont {Louie}}]{Spataru2005}%
  \BibitemOpen
  \bibfield  {author} {\bibinfo {author} {\bibfnamefont {C.~D.}\ \bibnamefont
  {Spataru}}, \bibinfo {author} {\bibfnamefont {S.}~\bibnamefont
  {Ismail-Beigi}}, \bibinfo {author} {\bibfnamefont {R.~B.}\ \bibnamefont
  {Capaz}}, \ and\ \bibinfo {author} {\bibfnamefont {S.~G.}\ \bibnamefont
  {Louie}},\ }\href {\doibase 10.1103/PhysRevLett.95.247402} {\bibfield
  {journal} {\bibinfo  {journal} {Phys. Rev. Lett.}\ }\textbf {\bibinfo
  {volume} {95}},\ \bibinfo {pages} {247402} (\bibinfo {year}
  {2005})}\BibitemShut {NoStop}%
\bibitem [{\citenamefont {Palummo}\ \emph {et~al.}(2015)\citenamefont
  {Palummo}, \citenamefont {Bernardi},\ and\ \citenamefont
  {Grossman}}]{Palummo2015}%
  \BibitemOpen
  \bibfield  {author} {\bibinfo {author} {\bibfnamefont {M.}~\bibnamefont
  {Palummo}}, \bibinfo {author} {\bibfnamefont {M.}~\bibnamefont {Bernardi}}, \
  and\ \bibinfo {author} {\bibfnamefont {J.~C.}\ \bibnamefont {Grossman}},\
  }\href {\doibase 10.1021/nl503799t} {\bibfield  {journal} {\bibinfo
  {journal} {Nano Letters}\ }\textbf {\bibinfo {volume} {15}},\ \bibinfo
  {pages} {2794} (\bibinfo {year} {2015})}\BibitemShut {NoStop}%
\bibitem [{\citenamefont {Lautenschlager}\ \emph {et~al.}(1987)\citenamefont
  {Lautenschlager}, \citenamefont {Garriga}, \citenamefont {Vina},\ and\
  \citenamefont {Cardona}}]{lautenschlager1987temperature}%
  \BibitemOpen
  \bibfield  {author} {\bibinfo {author} {\bibfnamefont {P.}~\bibnamefont
  {Lautenschlager}}, \bibinfo {author} {\bibfnamefont {M.}~\bibnamefont
  {Garriga}}, \bibinfo {author} {\bibfnamefont {L.}~\bibnamefont {Vina}}, \
  and\ \bibinfo {author} {\bibfnamefont {M.}~\bibnamefont {Cardona}},\
  }\href@noop {} {\bibfield  {journal} {\bibinfo  {journal} {Physical Review
  B}\ }\textbf {\bibinfo {volume} {36}},\ \bibinfo {pages} {4821} (\bibinfo
  {year} {1987})}\BibitemShut {NoStop}%
\bibitem [{\citenamefont {Roessler}\ and\ \citenamefont
  {Walker}(1967)}]{roessler1967optical}%
  \BibitemOpen
  \bibfield  {author} {\bibinfo {author} {\bibfnamefont {D.}~\bibnamefont
  {Roessler}}\ and\ \bibinfo {author} {\bibfnamefont {W.}~\bibnamefont
  {Walker}},\ }\href@noop {} {\bibfield  {journal} {\bibinfo  {journal} {JOSA}\
  }\textbf {\bibinfo {volume} {57}},\ \bibinfo {pages} {835} (\bibinfo {year}
  {1967})}\BibitemShut {NoStop}%
\bibitem [{\citenamefont {Haydock}(1980)}]{Haydock1980}%
  \BibitemOpen
  \bibfield  {author} {\bibinfo {author} {\bibfnamefont {R.}~\bibnamefont
  {Haydock}},\ }in\ \href@noop {} {\emph {\bibinfo {booktitle} {Solid State
  Physics, Advances in Research and Applications}}},\ Vol.~\bibinfo {volume}
  {35},\ \bibinfo {editor} {edited by\ \bibinfo {editor} {\bibfnamefont
  {F.}~\bibnamefont {Seitz}}\ and\ \bibinfo {editor} {\bibfnamefont
  {D.}~\bibnamefont {Turnbull}}}\ (\bibinfo  {publisher} {Academic New York},\
  \bibinfo {year} {1980})\BibitemShut {NoStop}%
\bibitem [{\citenamefont {Benedict}\ and\ \citenamefont
  {Shirley}(1999)}]{Benedict1999}%
  \BibitemOpen
  \bibfield  {author} {\bibinfo {author} {\bibfnamefont {L.~X.}\ \bibnamefont
  {Benedict}}\ and\ \bibinfo {author} {\bibfnamefont {E.~L.}\ \bibnamefont
  {Shirley}},\ }\href@noop {} {\bibfield  {journal} {\bibinfo  {journal} {Phys.
  Rev. B}\ }\textbf {\bibinfo {volume} {59}},\ \bibinfo {pages} {5441}
  (\bibinfo {year} {1999})}\BibitemShut {NoStop}%
\bibitem [{\citenamefont {Gr\"uning}\ \emph {et~al.}(2009)\citenamefont
  {Gr\"uning}, \citenamefont {Marini},\ and\ \citenamefont
  {Gonze}}]{Gruning2009}%
  \BibitemOpen
  \bibfield  {author} {\bibinfo {author} {\bibfnamefont {M.}~\bibnamefont
  {Gr\"uning}}, \bibinfo {author} {\bibfnamefont {A.}~\bibnamefont {Marini}}, \
  and\ \bibinfo {author} {\bibfnamefont {X.}~\bibnamefont {Gonze}},\
  }\href@noop {} {\bibfield  {journal} {\bibinfo  {journal} {Nano letters}\
  }\textbf {\bibinfo {volume} {9}},\ \bibinfo {pages} {2820} (\bibinfo {year}
  {2009})}\BibitemShut {NoStop}%
\bibitem [{\citenamefont {Bruneval}\ \emph {et~al.}(2006)\citenamefont
  {Bruneval}, \citenamefont {Vast},\ and\ \citenamefont
  {Reining}}]{Bruneval2006}%
  \BibitemOpen
  \bibfield  {author} {\bibinfo {author} {\bibfnamefont {F.}~\bibnamefont
  {Bruneval}}, \bibinfo {author} {\bibfnamefont {N.}~\bibnamefont {Vast}}, \
  and\ \bibinfo {author} {\bibfnamefont {L.}~\bibnamefont {Reining}},\
  }\href@noop {} {\bibfield  {journal} {\bibinfo  {journal} {Physical Review
  B}\ }\textbf {\bibinfo {volume} {74}},\ \bibinfo {pages} {045102} (\bibinfo
  {year} {2006})}\BibitemShut {NoStop}%
\bibitem [{\citenamefont {Gr\"uning}\ \emph {et~al.}(2016)\citenamefont
  {Gr\"uning}, \citenamefont {Sangalli},\ and\ \citenamefont
  {Attaccalite}}]{Gruning2016}%
  \BibitemOpen
  \bibfield  {author} {\bibinfo {author} {\bibfnamefont {M.}~\bibnamefont
  {Gr\"uning}}, \bibinfo {author} {\bibfnamefont {D.}~\bibnamefont {Sangalli}},
  \ and\ \bibinfo {author} {\bibfnamefont {C.}~\bibnamefont {Attaccalite}},\
  }\href {\doibase 10.1103/PhysRevB.94.035149} {\bibfield  {journal} {\bibinfo
  {journal} {Phys. Rev. B}\ }\textbf {\bibinfo {volume} {94}},\ \bibinfo
  {pages} {035149} (\bibinfo {year} {2016})}\BibitemShut {NoStop}%
\bibitem [{\citenamefont {Yabana}\ \emph {et~al.}(2012)\citenamefont {Yabana},
  \citenamefont {Sugiyama}, \citenamefont {Shinohara}, \citenamefont {Otobe},\
  and\ \citenamefont {Bertsch}}]{Yabana2012}%
  \BibitemOpen
  \bibfield  {author} {\bibinfo {author} {\bibfnamefont {K.}~\bibnamefont
  {Yabana}}, \bibinfo {author} {\bibfnamefont {T.}~\bibnamefont {Sugiyama}},
  \bibinfo {author} {\bibfnamefont {Y.}~\bibnamefont {Shinohara}}, \bibinfo
  {author} {\bibfnamefont {T.}~\bibnamefont {Otobe}}, \ and\ \bibinfo {author}
  {\bibfnamefont {G.~F.}\ \bibnamefont {Bertsch}},\ }\href {\doibase
  10.1103/PhysRevB.85.045134} {\bibfield  {journal} {\bibinfo  {journal} {Phys.
  Rev. B}\ }\textbf {\bibinfo {volume} {85}},\ \bibinfo {pages} {045134}
  (\bibinfo {year} {2012})}\BibitemShut {NoStop}%
\bibitem [{\citenamefont {Von~Neumann}(1927)}]{Neumann1927}%
  \BibitemOpen
  \bibfield  {author} {\bibinfo {author} {\bibfnamefont {J.}~\bibnamefont
  {Von~Neumann}},\ }\href@noop {} {\bibfield  {journal} {\bibinfo  {journal}
  {Nachrichten von der Gesellschaft der Wissenschaften zu G{\"o}ttingen,
  Mathematisch-Physikalische Klasse}\ }\textbf {\bibinfo {volume} {1927}},\
  \bibinfo {pages} {245} (\bibinfo {year} {1927})}\BibitemShut {NoStop}%
\bibitem [{Note16()}]{Note16}%
  \BibitemOpen
  \bibinfo {note} {This EOM can also be derived as an approximation to the
  non-equilibrium Green's function theory or quantum kinetic theory~\cite
  {kadanoff,Bonitz} where the density matrix is the time diagonal of the lesser
  Green function $\varrho (t)=-iG^<(t,t)$.}\BibitemShut {Stop}%
\bibitem [{\citenamefont {Tokman}(2009)}]{Tokman2009}%
  \BibitemOpen
  \bibfield  {author} {\bibinfo {author} {\bibfnamefont {M.}~\bibnamefont
  {Tokman}},\ }\href@noop {} {\bibfield  {journal} {\bibinfo  {journal}
  {Physical Review A}\ }\textbf {\bibinfo {volume} {79}},\ \bibinfo {pages}
  {053415} (\bibinfo {year} {2009})}\BibitemShut {NoStop}%
\bibitem [{\citenamefont {Gonze}\ \emph {et~al.}(1995)\citenamefont {Gonze},
  \citenamefont {Ghosez},\ and\ \citenamefont {Godby}}]{Gonze1995}%
  \BibitemOpen
  \bibfield  {author} {\bibinfo {author} {\bibfnamefont {X.}~\bibnamefont
  {Gonze}}, \bibinfo {author} {\bibfnamefont {P.}~\bibnamefont {Ghosez}}, \
  and\ \bibinfo {author} {\bibfnamefont {R.}~\bibnamefont {Godby}},\ }\href
  {\doibase 10.1103/PhysRevLett.74.4035} {\bibfield  {journal} {\bibinfo
  {journal} {Physical Review Letters}\ }\textbf {\bibinfo {volume} {74}},\
  \bibinfo {pages} {4035} (\bibinfo {year} {1995})}\BibitemShut {NoStop}%
\bibitem [{\citenamefont {Martin}\ and\ \citenamefont
  {Ortiz}(1997)}]{Martin1997}%
  \BibitemOpen
  \bibfield  {author} {\bibinfo {author} {\bibfnamefont {R.~M.}\ \bibnamefont
  {Martin}}\ and\ \bibinfo {author} {\bibfnamefont {G.}~\bibnamefont {Ortiz}},\
  }\href@noop {} {\bibfield  {journal} {\bibinfo  {journal} {Physical Review
  B}\ }\textbf {\bibinfo {volume} {56}},\ \bibinfo {pages} {1124} (\bibinfo
  {year} {1997})}\BibitemShut {NoStop}%
\bibitem [{Note17()}]{Note17}%
  \BibitemOpen
  \bibinfo {note} {For a discussion and a general expression, valid at all
  orders, see Ref.~\protect \rev@citealpnum {Attaccalite2013}}\BibitemShut
  {NoStop}%
\bibitem [{Note18()}]{Note18}%
  \BibitemOpen
  \bibinfo {note} {While the density-density response function can be used only
  to compute the longitudinal response, the dipole-dipole response function,
  like the current-current one, can also describe transverse and mixed
  (longitudinal-transverse) terms.}\BibitemShut {Stop}%
\bibitem [{\citenamefont {Nozi{\`e}res}\ and\ \citenamefont
  {Pines}(1999)}]{NozieresPines}%
  \BibitemOpen
  \bibfield  {author} {\bibinfo {author} {\bibfnamefont {P.}~\bibnamefont
  {Nozi{\`e}res}}\ and\ \bibinfo {author} {\bibfnamefont {D.}~\bibnamefont
  {Pines}},\ }\href@noop {} {\emph {\bibinfo {title} {Theory of quantum
  liquids}}}\ (\bibinfo  {publisher} {Westview Press},\ \bibinfo {year}
  {1999})\BibitemShut {NoStop}%
\bibitem [{\citenamefont {Adler}(1962)}]{Adler1962}%
  \BibitemOpen
  \bibfield  {author} {\bibinfo {author} {\bibfnamefont {S.~L.}\ \bibnamefont
  {Adler}},\ }\href@noop {} {\bibfield  {journal} {\bibinfo  {journal} {Phys.
  Rev.}\ }\textbf {\bibinfo {volume} {126}},\ \bibinfo {pages} {413} (\bibinfo
  {year} {1962})}\BibitemShut {NoStop}%
\bibitem [{Note19()}]{Note19}%
  \BibitemOpen
  \bibinfo {note} {The direction of the ${\protect \bf q}\rightarrow {\protect
  \bf 0}$ limit is a delicate point. Indeed the $({\protect \bf 0},0)$ point in
  the $({\protect \bf q},\omega )$ plane is non analytic, i.e. the value of the
  dielectric function depends on the direction of the limit $({\protect \bf
  q},\omega )\rightarrow ({\protect \bf 0},0)$. This direction is determined by
  the experiment we would like to describe. In optical experiments the
  direction of interest is $\omega =c{\protect \bf q}$. Other directions in the
  $({\protect \bf q},\omega )$ plane may be of interest. For example in
  electron-energy loss experiments one measures the inverse dielectric function
  $\varepsilon ^{-1}({\protect \bf q},\omega )$ at fixed momentum ${\protect
  \bf q}_{exp}$. The line ${\protect \bf q}={\protect \bf q}_{exp}$ always
  crosses the $\varv ^F({\protect \bf q})$ line, thus a peak must always appear
  if we derive $\varepsilon ({\protect \bf q},\omega )$ from the computed
  electron energy loss function at ${\protect \bf q}_{exp}$. Therefore for
  ${\protect \bf q}_{exp}\rightarrow {\protect \bf 0}$ we always obtain the
  Drude-like tail in the absorption spectrum, also without
  smearing.}\BibitemShut {Stop}%
\bibitem [{\citenamefont {Sangalli}\ \emph {et~al.}(2016)\citenamefont
  {Sangalli}, \citenamefont {Dal~Conte}, \citenamefont {Manzoni}, \citenamefont
  {Cerullo},\ and\ \citenamefont {Marini}}]{Sangalli2016}%
  \BibitemOpen
  \bibfield  {author} {\bibinfo {author} {\bibfnamefont {D.}~\bibnamefont
  {Sangalli}}, \bibinfo {author} {\bibfnamefont {S.}~\bibnamefont {Dal~Conte}},
  \bibinfo {author} {\bibfnamefont {C.}~\bibnamefont {Manzoni}}, \bibinfo
  {author} {\bibfnamefont {G.}~\bibnamefont {Cerullo}}, \ and\ \bibinfo
  {author} {\bibfnamefont {A.}~\bibnamefont {Marini}},\ }\href {\doibase
  10.1103/PhysRevB.93.195205} {\bibfield  {journal} {\bibinfo  {journal} {Phys.
  Rev. B}\ }\textbf {\bibinfo {volume} {93}},\ \bibinfo {pages} {195205}
  (\bibinfo {year} {2016})}\BibitemShut {NoStop}%
\bibitem [{\citenamefont {Blount}(1962)}]{Blount1962}%
  \BibitemOpen
  \bibfield  {author} {\bibinfo {author} {\bibfnamefont {E.}~\bibnamefont
  {Blount}},\ }in\ \href@noop {} {\emph {\bibinfo {booktitle} {Solid State
  Physics, Advances in Research and Applications}}},\ \bibinfo {editor} {edited
  by\ \bibinfo {editor} {\bibfnamefont {F.}~\bibnamefont {Seitz}}\ and\
  \bibinfo {editor} {\bibfnamefont {D.}~\bibnamefont {Turnbull}}}\ (\bibinfo
  {publisher} {Academinc New York},\ \bibinfo {year} {1962})\BibitemShut
  {NoStop}%
\bibitem [{\citenamefont {Souza}\ \emph {et~al.}(2004)\citenamefont {Souza},
  \citenamefont {\'I\~niguez},\ and\ \citenamefont {Vanderbilt}}]{Souza2004}%
  \BibitemOpen
  \bibfield  {author} {\bibinfo {author} {\bibfnamefont {I.}~\bibnamefont
  {Souza}}, \bibinfo {author} {\bibfnamefont {J.}~\bibnamefont {\'I\~niguez}},
  \ and\ \bibinfo {author} {\bibfnamefont {D.}~\bibnamefont {Vanderbilt}},\
  }\href {\doibase 10.1103/PhysRevB.69.085106} {\bibfield  {journal} {\bibinfo
  {journal} {Phys. Rev. B}\ }\textbf {\bibinfo {volume} {69}},\ \bibinfo
  {pages} {085106} (\bibinfo {year} {2004})}\BibitemShut {NoStop}%
\bibitem [{\citenamefont {Kadanoff}\ and\ \citenamefont
  {Baym}(1962)}]{kadanoff}%
  \BibitemOpen
  \bibfield  {author} {\bibinfo {author} {\bibfnamefont {L.~P.}\ \bibnamefont
  {Kadanoff}}\ and\ \bibinfo {author} {\bibfnamefont {G.~A.}\ \bibnamefont
  {Baym}},\ }\href@noop {} {\emph {\bibinfo {title} {Quantum statistical
  mechanics}}}\ (\bibinfo  {publisher} {Benjamin},\ \bibinfo {year}
  {1962})\BibitemShut {NoStop}%
\bibitem [{\citenamefont {Bonitz}(1998)}]{Bonitz}%
  \BibitemOpen
  \bibfield  {author} {\bibinfo {author} {\bibfnamefont {M.}~\bibnamefont
  {Bonitz}},\ }\href@noop {} {\emph {\bibinfo {title} {Quantum Kinetic
  Theory}}}\ (\bibinfo  {publisher} {Yeubner-Verlag Stuttgart/Leipzig},\
  \bibinfo {year} {1998})\BibitemShut {NoStop}%
\bibitem [{\citenamefont {Attaccalite}\ and\ \citenamefont
  {Gr{\"u}ning}(2013)}]{Attaccalite2013}%
  \BibitemOpen
  \bibfield  {author} {\bibinfo {author} {\bibfnamefont {C.}~\bibnamefont
  {Attaccalite}}\ and\ \bibinfo {author} {\bibfnamefont {M.}~\bibnamefont
  {Gr{\"u}ning}},\ }\href@noop {} {\bibfield  {journal} {\bibinfo  {journal}
  {Physical Review B}\ }\textbf {\bibinfo {volume} {88}},\ \bibinfo {pages}
  {235113} (\bibinfo {year} {2013})}\BibitemShut {NoStop}%
\end{thebibliography}%
\end{document}